# The Evolution of the Star-forming Interstellar Medium across Cosmic Time

**Linda J. Tacconi[1], Reinhard Genzel[1,2] & Amiel Sternberg[3,4]**

[1]*Max-Planck-Institut für extraterrestrische Physik (MPE), Giessenbachstr.1, 85748 Garching, Germany*

[2]*Departments of Physics and Astronomy, University of California, 94720 Berkeley, USA*

[3]*School of Physics and Astronomy, Tel Aviv University, Tel Aviv 69978, Israel*

[4] *Center for Computational Astrophysics, Flatiron Institute, 162 5th Avenue, New York, NY 10010, USA*

# Abstract

Over the past decade increasingly robust estimates of the dense molecular gas content in galaxy populations between redshift z=0 and the peak of cosmic galaxy/star formation (z~1-3) have become available. This rapid progress has been possible due to the advent of powerful ground-based, and space telescopes for combined study of several millimeter to far-IR, line or continuum tracers of the molecular gas and dust components. The main conclusions of this review are:

- Star forming galaxies contained much more molecular gas at earlier cosmic epochs than at the present time.

- The galaxy integrated depletion time scale for converting the gas into stars depends primarily on z or Hubble time, and at a given z, on the vertical location of a galaxy along the star-formation rate versus stellar mass "main-sequence" (MS) correlation.

- Global rates of galaxy gas accretion primarily control the evolution of the cold molecular gas content and star formation rates of the dominant MS galaxy population, which in turn vary with the cosmological expansion. A second key driver may be global disk fragmentation in high-z, gas rich galaxies, which ties local free-fall time scales to galactic orbital times, and leads to rapid radial matter transport and bulge growth. Third, the low star formation efficiency inside molecular clouds is plausibly set by super-sonic streaming motions, and internal turbulence, which in turn may be driven by conversion of gravitational energy at high-z, and/or by local feedback from massive stars at low-z.

- A simple 'gas regulator' model is remarkably successful in predicting the combined evolution of molecular gas fractions, star formation rates, galactic winds, and gas phase metallicities.

# Table of Contents

# 1. **Introduction**

It has been over 40 years since the first detections of molecular gas in galaxies outside the Milky Way were published (Rickard et al. 1975, 1977; Solomon & de Zafra 1975; see also Young & Scoville 1991). Since then it has become possible to map the distribution and kinematics of molecular gas in external galaxies with powerful (sub)millimeter interferometer arrays. The first high-redshift (high-$z$) CO detections in very luminous active galactic nuclei (AGNs) and submillimeter galaxies (SMGs) were reported by Scoville et al. (1995, 1997) and Frayer et al (1998, 1999), and the first detections of CO in 'normal' high-$z$ main-sequence (MS) star-forming galaxies (SFGs) were in Tacconi et al. (2010) and Daddi et al. (2010). Molecular gas[1] and dust studies of a few thousand normal SFGs from redshifts $z$=0 to 4 have been published in the last decade. The frontier is now at $z>7$. Earlier concerns about the applicability of the 'CO-to-$H_2$ conversion factor' have now been largely allayed with the availability of dust-based estimates of molecular gas masses, at least on galaxy integrated scales. While there have been several recent reviews summarizing molecular gas studies (e.g. Carilli and Walter 2013; Combes 2018) in this article we strive to place the studies of cold dense gas within the framework of cosmological galaxy formation and evolution. We begin by summarizing the properties of galaxy populations across cosmic time. We then show how knowledge of the cold gas contents has provided critical new information on galaxy evolution. As we progress, we emphasize the key physical processes, and compare the observations to semi-analytic models and hydro-simulations (c.f. Somerville & Davé 2015). We conclude with questions we believe are next on the agenda for substantial progress.

## 1.1 Star Formation in Molecular Clouds

Stars form from dusty, molecular interstellar gas (McKee & Ostriker 2007; Kennicutt & Evans 2012). In the Milky Way and nearby galaxies all star formation appears to occur in massive ($M\sim10^4\ldots10^{6.5}$ $M_\odot$), dense ($n(H_2)\sim10^2\ldots10^5$ cm$^{-3}$) and cold ($T_{gas}\sim$10-40 K) '(giant) molecular clouds' ((G)MCs) of diameter $2R\sim$50-100 pc (Solomon et al. 1987; Bolatto et al. 2008; McKee & Ostriker 2007). GMCs are highly supersonic ($\sigma\sim$0.7 km/s $\times(R$/parsec$)^{0.5}$) with Mach numbers ~30 (Larson 1981; Solomon et al. 1987; Elmegreen & Scalo 2004; Scalo & Elmegreen 2004). They are near but somewhat above virial equilibrium (that is, unbound), with virial parameters $\alpha_V=5\sigma^2R/GM$ between 1 and 10, and a median of ~2 (Miville-Deschenes 2017; Sun et al. 2018; Meidt et al. 2018; Schruba, Kruijssen & Leroy 2019). On scales of individual star formation regions, clouds are likely transient with GMC life-times of $t_{GMC}\sim$ 5-20 Myr, due to disruption by protostellar outflows, ionized gas flows and internal supernova-explosions (e.g. Elmegreen 2007; Ballestero-Paredes 2007; Dobbs et al. 2011; Dobbs & Pringle 2013; Kruijssen et al. 2019; Chevance et al. 2019).

---

[1] throughout this paper we correct $H_2$ masses upward by 1.36 for the content for helium and heavy elements for a census of the entire mass content of the molecular phase.

Star formation rates (SFRs) on galactic scales, or star formation surface densities on sub-galactic scales down to a few kpc, are most strongly correlated with molecular gas (or dust) masses, or surface densities. There is little or no correlation between star formation and neutral atomic hydrogen at low surface densities (Kennicutt 1989; Wong & Blitz 2002; Kennicutt et al. 2007; Bigiel et al. 2008, 2011; Leroy et al. 2008, 2013; Schruba et al. 2011). It is not clear whether high molecular content is causally required for the onset of star formation (Glover & Clark 2012). More likely high gas volume densities and sufficient dust shielding ($A_V$>7, $\Sigma_{gas}$>100 $M_\odot pc^{-2}$) decouple the dense cores from the external radiation fields, allowing the clouds to cool rapidly and collapse. These conditions may then also be naturally conducive to molecule formation, which enhances further cooling (Sternberg & Dalgarno 1989; Glover & Clark 2012; Krumholz, Leroy & McKee 2011; Heiderman et al. 2010; Lada et al. 2012).

A self-gravitating molecular gas cloud of mean molecular hydrogen density $\rho_{H2}$ has a local free fall time $t_{ff} = \sqrt{3\pi/(32G\rho_{H2})}$. Heuristically, above a threshold of ~10 $M_\odot pc^{-2}$ (section 4.5), a self-gravitating cloud of total mass M should then form stars at a rate $SFR = \varepsilon_{ff} \times \frac{M}{t_{ff}}$. Here $\varepsilon_{ff}$ is the efficiency of star formation per free-fall time (c.f. Kennicutt 1998; Elmegreen 2002; Krumholz & McKee 2005). For a homogeneous cloud supported by thermal gas motions, this efficiency (per free fall time) should be high. For example, Alves et al. (2007), André et al. (2010) and Könyves et al. (2015) find a clump to star conversion efficiency of 20-40% in thermal cloud cores. However, ever since the first observations of molecular gas and cloud collapse in the Milky Way (MW) it has become clear that on larger cloud and galactic scales the efficiency is small, $\varepsilon_{ff}$~ one to a few percent (Zuckerman & Evans 1974; Krumholz & Tan 2007).

There are two commonly accepted explanations for this low efficiency. One is that magnetic fields pervade the GMCs, which are 'frozen in' because of trace ions created by cosmic ray ionization. These fields stabilize the initially magnetically subcritical clouds against gravitational collapse (e.g. Mouschovias 1976; Shu et al. 1987). Ambipolar diffusion on a time scale of ~10 $t_{ff}$ then allows a gradual cloud collapse of the cloud. . Observations suggest that cloud cores are magnetically super-critical, and thus cannot be stabilized by magnetic pressure (Crutcher 2012). The second explanation rests on turbulent support. The highly supersonic gas motions in GMCs, and the interplay of dispersive and compressive shocks, prevent most of the gas from collapsing at any given time (Stone et al. 1998; Mac Low 1999; Klessen et al. 2000; Elmegreen & Scalo 2004; Krumholz & McKee 2005; McKee & Ostriker 2007). Since super-sonic motions are dispersed on a dynamical time, the turbulent energy must be continuously replenished, either within clouds by outflows and super-novae explosions, or by external gravitational torques or shear motions (Stone et al. 1998; McKee & Ostriker 2007).

## 1.2 Galaxy and Star Formation across Cosmic Time

The formation/evolution of galaxies across cosmic time[2] is complex. It involves the hierarchical merging of virialized dark matter halos, the accretion and cooling of gas onto the growing galaxies, the formation of stars in cold dense gas clouds and outflows of heavy-element enriched gas into the circum-galactic medium (CGM), driven by massive stars, supernovae, and accreting supermassive black holes in the galaxy nuclei (e.g. Rees & Ostriker 1977; White & Rees 1978; White & Frenk 1991; Kauffmann et al. 1993; Croton et al. 2006; Bournaud & Elmegreen 2009; Dekel et al. 2009; Dekel & Mandelker 2014; Bouché et al. 2010; Guo et al. 2011; Davé et al. 2011, 2012; Davé, Oppenheimer & Finlator 2011a,b, 2012; Lilly et al. 2013). Multi-wavelength observations over the past two decades have provided an increasingly detailed picture of this '**baryonic cycle**'. Comprehensive studies show that the star-formation rate density peaked approximately 3.5 Gyr after the Big Bang, at $z \approx 1.5$-$2.5$, and has declined since by a factor 10-15 to the present epoch, with an e-folding timescale of 4 Gyr (Lilly et al. 1996; Madau et al. 1996; Steidel et al. 1996; Madau & Dickinson 2014, hereafter MD14)[3]. The cosmic star formation history and stellar mass growth in galaxies inferred from the measurements are depicted as filled red circles in the left and right panels of Figure 1 (adopted from MD14). The first galaxies condensed at z~8-11, 350-550 Myrs after the Big Bang (e.g. Bouwens et al. 2014; Oesch et al. 2016). Half the stellar mass observed today was formed prior to z = 1.3. Assuming a universal stellar initial mass function (IMF), the global stellar mass density at any epoch matches reasonably well the time integral of all the preceding star-formation activity (right panel of Figure 1). The co-moving rates of star formation and central black hole accretion follow a similar rise and fall, offering evidence for a sustained co-evolution of black holes and their host galaxies (MD14, left panel of Figure 1). As we will show in sections 3 and 4 the co-moving molecular gas densities and molecular gas fractions show a similar rise and fall, although these quantities are still very uncertain for z>3.

## 1.3 The Star Formation Main Sequence

About 90% of the cosmic star formation between z=0 and 2.5 occurs in galaxies that lie along the so-called '**star formation main sequence**', or '**MS**' (Rodighiero et al. 2011, 2015). The MS is a fairly tight (±0.3 dex scatter), **near-linear relationship between stellar mass and star formation rate** (Brinchmann et al. 2004; Schiminovich et al. 2007; Noeske et al. 2007; Elbaz et al. 2007, 2011; Daddi et al. 2007; Peng et al. 2010; Rodighiero et al. 2010, 2011; Karim et al. 2011; Whitaker et al. 2012, 2014; Renzini & Peng 2015; Speagle et al. 2014, hereafter S14; Schreiber et al. 2015).

For galaxies on the MS the dependence of the specific star formation rate, $sSFR=SFR/M_*$, on stellar mass varies slowly with stellar mass as $sSFR \sim M_*^{-0.1..-0.4}$, but the zero point increases very strongly with redshift, *sSFR* ∝ $(1+z)^3$ up to $z\sim2$, and ∝ $(1+z)^{1.5}$ for $z>2$ (e.g. Lilly et al. 2013). Several studies have captured these dependencies through empirical fittings to different galaxy samples from deep surveys (e.g. Whitaker et al. 2012, 2014, S14, amongst others). The results are mostly in good agreement, with differences in zero-points and slopes in the mass and redshift relations depending on sample selections (redshift range, survey bands), survey completeness, and methodologies applied to derive $M_*$ and *SFR*s (Renzini & Peng 2015).

---

[2] throughout this review we use flat, ΛCDM cosmology with $H_0$=70 km/s Mpc$^{-1}$ and $\Omega_m$=0.3.
[3] unless stated otherwise we adopt a Chabrier (2003) or Kroupa (2001) initial stellar mass function, and have changed input values from the literature accordingly

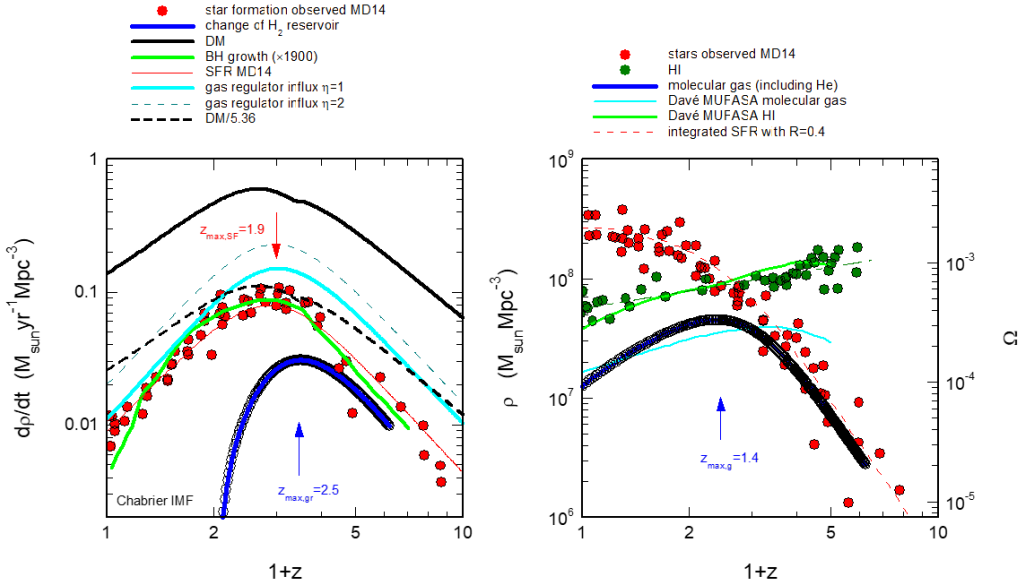

*Figure 1. Cosmic evolutions of gas, stars and massive black holes. Left panel: Evolution of cosmic star formation history per co-moving cosmic volume (red filled circles and red fit function line), from a wide range of UV- to far-IR multi-band surveys (adopted from MD14, their Figure 9c). For comparison we show the rates of dark matter accretion (thick black), dark matter accretion divided by the cosmic DM/baryon ratio of 5.36 (dashed black), and the growth rate of massive black holes (green, Figure 15 of MD14). The continuous and dotted cyan curves denote the baryon gas accretion rates required by the 'gas regulator' model (Bouché et al. 2010, Lilly et al. 2013) for a wind mass loading factor of 1 and 2, respectively. Finally, the thick blue/black curve is the rate of change of the molecular gas reservoir in galaxies estimated from this model (section 4.3). Right panel: Red circles denote the stellar mass in galaxies as observed in multi-band estimates (Figure 11 of MD14) and the continuous red line denotes the mass inferred from integrating the SFRs in the left panel, for a Chabrier (2003) IMF and a return fraction R of 40%. The thick blue-black curve shows the evolution of the molecular mass (including He) reservoir. Dark green filled circles and the connecting power law fit (slope 0.5) denote the evolution of atomic gas (adapted from Figure 14 in Rhee et al. 2017 and references therein and in the text). The light green and cyan curves denote the HI and molecular gas ($H_2*1.36$) content obtained in the MUFASA simulation of Davé et al. (2017).*

Excepting for passive galaxies at late times, or galaxies in dense environments, star forming field galaxies mainly grew in mass along the MS through star formation (Guo & White 2008). The stellar mass function at all z has an exponential cutoff above the Schechter mass, $M_S \sim 10^{10.7-10.9}$ $M_\odot$ (Peng et al. 2010; Ilbert et al. 2010, 2013). Above this mass, star formation appears to quench (see sidebar).

Figure 2 shows the MS-lines derived by S14, and Whitaker et al. (2012, 2014), shown as *log(sSFR)* vs *log(1+z)*, corrected to a fiducial stellar mass of $5 \times 10^{10}$ $M_\odot$ and log*sSFR* vs log$M_*$, corrected to a common redshift of *z*=1.5. A crucial point is that the results differ dramatically depending on whether one infers *SFRs* from UV plus mid- or far-infrared (24μm, 70-160μm) photometry or from SED fits; the UV+IR method generally better accounts for dust extinction and results in higher SFR estimates (e.g. Wuyts et al. 2011a). In practice the UV+IR method is not available for all galaxies at *z*>3, nor for galaxies below the main sequence at *z*>1.5. In this review, we adopt *SFRs* from UV+IR methodology where possible, and from SED-synthesis modeling for other cases – a strategy sometimes referred to as the 'ladder technique' (Wuyts et al. 2011a).

In Figure 2 and for the rest of the paper (see also Tacconi et al. 2018, hereafter T18), we adopt the main sequence definition of S14,

$$\log(sSFR(MS,z,M_*) \; (\text{Gyr}^{-1})) = (-0.16 - 0.026 \times t_c(Gyr)) \times (\log M_*(M_\odot) + 0.025)$$
$$- (6.51 - 0.11 \times t_c(Gyr)) + 9 \; ,$$
$$\text{with} \quad \log t_c(Gyr) = 1.143 - 1.026 \times \log(1+z) - 0.599 \times \log^2(1+z)$$
$$+ 0.528 \times \log^3(1+z) \qquad (1).$$

Here, $t_c$ (Gyr) is the cosmic time[2]. The S14 prescription is applicable over a wide range of redshifts from z=0-5, and a wide range of stellar masses from log($M_*/M_\odot$)=9.0-11.8, optimal for cold gas evolution comparisons. The main disadvantage of S14 is that it is calibrated mostly on SED-based *SFRs*, which as discussed above, tend to be lower than UV-IR based *SFRs*. Except for those issues, the S14 formula is very similar to the relations proposed by Whitaker et al. (2012, 2014).

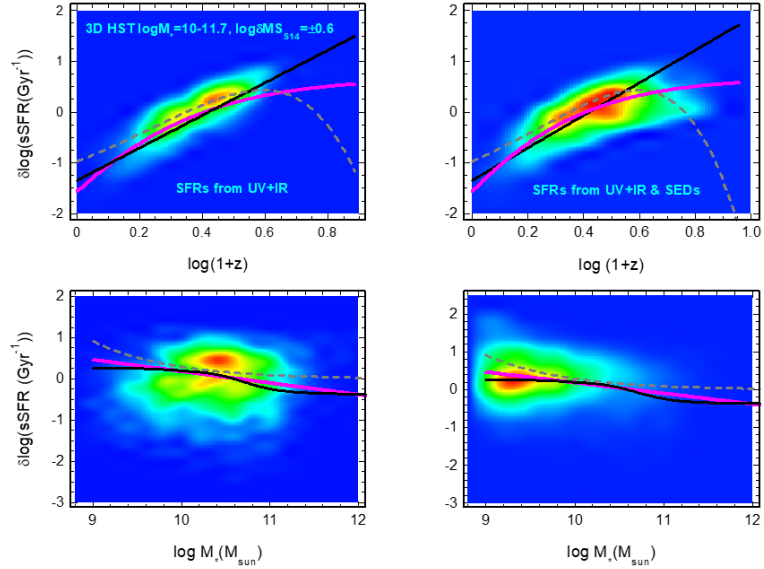

***Figure 2:*** *Specific star formation rates, sSFR=SFR/M$_*$ (Gyr$^{-1}$) as a function of log (1+z) at log(M$_*$/M$_☉$)=10.7 (top panels) and as a function of logM$_*$ for z=1.5 (bottom panels). The color distributions represent the distribution of galaxies in the 3D-HST survey on a linear scaling (Brammer et al. 2012, Skelton et al. 2014, Momcheva et al. 2016). In the left panels we show 3D-HST galaxies with log(M$_*$/M$_☉$)=10-11.7, logδMS=±0.6, which have individual 24μm Spitzer, or 70 μm, 100μm or 160μm Herschel detections, so that a IR+UV luminosity can be computed (Wuyts et al. 2011a). The right panels in addition include galaxies between log(M$_*$/M$_☉$)=9-10 and galaxies across the entire mass range with only an SED-based SFR, typically resulting in underestimated SFRs, which is particularly relevant at high-z, low log(M$_*$) and below the MS. We used the S14 MS prescription (Equation (1) in the main text) to correct all galaxies to the same mass of log(M$_*$/M$_☉$)=10.7 in the left plot, and to the same redshift (z=1.5, t$_c$=4.7 Gyr) in the right plot. The solid magenta, dotted grey and solid black lines denote the S14, Whitaker et al. (2012) and Whitaker et al. (2014) prescriptions of the MS, respectively. For this paper, we use S14 as our default prescription (Equation (1), adopted from T18).*

## 1.4 Structure of Main Sequence Galaxies

At all redshifts up to the (z~2) peak of the galaxy/star formation rate, main sequence SFGs have disky, exponential rest-frame optical light distributions with Sersic indices, $n_{Sersic}$~1-2 (Wuyts et al. 2011b); this despite the often clumpy and irregular appearance of *z*>1 SFGs in the rest-frame UV. The majority of massive (log*M$_*$*/M$_☉$)>10) SFGs are rotationally supported disks (e.g. Wisnioski et al. 2015, 2019, Simons et al. 2017). With increasing redshift, the fraction of lower mass SFGs with 'dispersion' dominated kinematics increases, suggesting that these systems are not settled, equilibrium disks (Kassin et at . 2012, Newman et al. 2013, Simons et al. 2017, Wisnioski et al. 2019). There are also sub-galactic, random motions due to unresolved streaming and a floor of galaxy-wide local 'turbulence' (Elmegreen & Scalo 2004; Scalo & Elmegreen 2004). In star forming ionized gas, the 1D rms velocity dispersion increases with redshift as $\sigma_0$ (km/s) ~ a + b×$(1+z)$ (a~b~10-11 km/s, (Übler et al. 2019; see also Förster Schreiber et al. 2006, 2009; Kassin et al. 2007, 2012; Shapiro et al. 2008; Epinat et al. 2012; Newman et al. 2013; Stott et al. 2016; Simons et al. 2017; Wisnioski et al. 2015; Bezanson et al. 2018). The z=0 value of the velocity dispersion in molecular gas is smaller (a~0, $\sigma_0(z=0)_{mol}$~10 km/s but the slope with redshift is the same as for the ionized gas (Übler et al. 2019), so that at high-z ionized and molecular gas dispersions are broadly comparable. The tightness and constant shape of the MS suggest that at any cosmic epoch star forming galaxies grow along the sequence in an equilibrium of gas accretion, star formation and gas outflows (Bouché et al. 2010; Davé, Finlator & Oppenheimer 2012; Lilly et al. 2013; Peng & Maiolino 2014; see also section 4.3).

Many recent studies have aimed at understanding the origin and evolution of the MS by focussing on molecular gas studies of galaxies at different redshifts (e.g. Daddi et al. 2010a,b; Tacconi et al. 2010, 2013, T18; Genzel et al. 2010, 2015 (hereafter G15); Bouché et al. 2010; Lilly et al. 2013; Davé et al. 2011a,b, 2012; Lagos et al. 2011, 2012, 2015; Fu et al. 2012; Scoville et al. 2016, 2017, hereafter S17; DeCarli et al. 2016, 2019). For more details see sections 2-4 of this review.

# 2. Estimating the Cold Mass Content of Galaxies

## 2.1 The CO Line Luminosity Method

Observations of GMCs in the Milky Way and nearby galaxies have established that the integrated flux of $^{12}$CO millimeter rotational lines can be used to infer molecular gas masses, although the CO molecule only makes up a small fraction (~$10^{-4}$) of the entire gas mass, and its lower rotational lines (1-0, 2-1, 3-2) are almost always very optically thick ($\tau_{CO}$~10; Dickman et al. 1986; Solomon et al. 1987; Bolatto et al. 2013). This is because the CO emission arises in moderately dense (volume densities $<n(H_2)>$ ~200 cm$^{-3}$, column densities $N(H_2)$~$10^{22}$ cm$^{-2}$), GMCs of kinetic temperature 10-50 K. Dickman et al. (1986) and Solomon et al. (1987) have shown that in this 'virial' regime, or when the emission comes from a cloud ensemble with similar mass and size and spread in velocity by galactic rotation ('cloud counting'), the integrated line CO line luminosity $L'_{CO} = \int_{source} \int_{line} T_R \, dv \, dA$

(in K km/s pc$^2$) is proportional to the total gas mass in the cloud or galaxy. Here, $T_R$ is the Rayleigh-Jeans source brightness temperature as a function of Doppler velocity $v$. In this regime the total molecular gas mass (including a 36% mass correction for helium) depends on the observed CO J →J-1 line flux $F_{CO\,J}$, source luminosity distance $D_L$, redshift $z$ and observed line wavelength $\lambda_{obs\,J} = \lambda_{rest\,J}\,(1+z)$ as (Solomon et al. 1997),

$$M_{molgas}/M_{sun} = \alpha_{CO\ 1} \times L'_{CO\ 1}$$

$$= 1.58 \times 10^9 \left(\frac{\alpha_{CO\ 1} \times R_{1J}}{\alpha_0}\right) \times \left(\frac{F_{CO\ J}}{\text{Jy km/s}}\right) \times (1+z)^{-3} \times \left(\frac{\lambda_{obs\ J}}{\text{mm}}\right)^2 \times \left(\frac{D_L}{\text{Gpc}}\right)^2 \quad (2).$$

Here $\alpha_{CO\ 1}$ is the empirical 'conversion factor' to transform the observed quantity (CO luminosity in the 1-0 transition) to the inferred physical quantity (molecular gas mass), and $R_{1J}$ is the ratio of the 1-0 to the $J \rightarrow J-1$ CO line luminosity, $R_{1J}=L'_{CO\ 1-0}/L'_{CO\ J-(J-1)}$.

*2.1.1 From Conversion 'Factor' to Conversion 'Function'.* The CO conversion factor is expected to depend on several physical parameters (Narayanan et al. 2011, 2012; Feldmann et al. 2012 a, b). In the virial/cloud counting model α depends on the ratio of the square root of the average cloud density $<n(H_2)>$ and the equivalent Rayleigh-Jeans brightness temperature $T_{R\ J}$ of the CO transition $J \rightarrow J-1$. Because of photo-dissociation of CO by UV photons in the outer cloud layers, it also increases inversely with metallicity Z (c.f. Leroy et al. 2011; Genzel et al. 2012; Bolatto et al. 2013), such that

$$\alpha_{CO\ J} = \alpha_0 \times \zeta\left(\frac{(<n(H_2)>)^{1/2}}{T_{R\ J}}\right) \times \chi(Z)$$ . The functions $\chi$ and $\zeta$ will be discussed in 2.1.2

and 2.1.4. In the Milky Way, nearby SFGs with near solar metallicity, and in dense star forming clumps of lower mass, lower metallicity galaxies, the empirical CO 1-0 conversion factor $\alpha_{CO\ 1}$ has been determined through dynamical, dust and γ-ray calibrations (see Bolatto et al. 2013 for a review). These are broadly consistent with a single value of $\alpha_{CO1} = \alpha_0 = 4.36 \pm 0.9$ ($M_\odot$/(K km/s pc$^2$)), equivalent to $X_{CO}=N(H_2)/(T_{RJ=1}\Delta v)= 2\times10^{20}$ (cm$^{-2}$/(K km/s)) (Bolatto et al. 2013; Strong & Mattox 1996; Dame et al. 2001; Grenier et al. 2005; Bolatto et al. 2008; Leroy et al. 2011; Abdo et al. 2010; Ostriker et al. 2010). Scoville et al. (2014, 2016) and S17 advocate for $\alpha_0=6.5$ (50% larger than the value above), based on a virial analysis of Milky Way GMCs. **Throughout this review we adopt** $\alpha_0 = 4.36$, consistent also with Daddi et al. (2010a,b, who adopt $\alpha_{CO}=3.6\pm0.8$, which includes HI and Helium).

*2.1.2 Metallicity Dependence of the Conversion Factor.* For galaxies with sub-solar gas phase metallicity, the conversion factor and metallicity are inversely correlated because CO is photo-dissociated (and the atomic carbon is photo-ionized) in an increasing fraction of the molecular hydrogen gas column. The result is that the H$_2$ gas is deficient ('dark') in CO (Wilson 1995; Arimoto et al. 1996; Israel 2000; Wolfire et al. 2010; Leroy et al. 2011; Genzel et al. 2012; Bolatto et al. 2013; Nordon & Sternberg 2016). Motivated by the theoretical work on CO photo-dissociation in clouds with a range of hydrogen densities and UV radiation field intensities, but with a constant hydrogen column (Wolfire et al. 2010), Bolatto et al. (2013) proposed $\chi(Z) = 0.67 \times \exp(0.36 \times 10^{-(12+\log(O/H)-8.67)})$ . Here $Z=12+\log(O/H)$ is the gas phase oxygen abundance in the galaxy on the Pettini & Pagel (2004) calibration scale, with the solar abundance of $Z_\odot=8.67$ (Asplund et al. 2004). The relation assumes an average GMC hydrogen column density of $9\times10^{21}$ cm$^{-2}$ or mass surface density of 100 $M_\odot$ pc$^{-2}$. Genzel et al. (2012) combined local (Leroy et al. 2011) and high-z empirical evidence to derive a second fitting function, $\chi(Z) = 10^{-1.27\times(12+\log(O/H)-8.67)}$. For metallicities >0.5×$Z_\odot$ the two fitting functions yield values within ±0.2 dex of each other but the former increases more strongly at low Z. Other studies found similar metallicity dependencies (Israel 1997; Wolfire et al. 2010; Schruba et al. 2011; Feldmann et al. 2012a, b; Sargent et al. 2014).

Bisbas et al. (2015) have pointed out that in galaxies with enhanced cosmic-ray rates (which would be correlated with enhanced star formation rates) CO can also be destroyed deep within clouds (see Bialy & Sternberg 2015), and might significantly affect the spatial distribution of CO in a galaxy, and thus, effectively, the conversion factor.

*2.1.3 CO Ladder Excitation Dependence of the Conversion Factor.* Observations of CO rotational ladders in near-MS galaxies between z=0 and 3 yield median correction factors of $R_{JI}$ = 1.3, 1.8 and 2.4 for the 2-1, 3-2 and 4-3 transitions (Weiss et al. 2007; Dannerbauer et al. 2009; Ivison et al. 2011; Riechers et al. 2010; Combes et al. 2013; Bauermeister et al. 2013; Bothwell et al. 2013; Aravena et al. 2014; Daddi et al. 2015). This 'excitation' correction entails a combination of the Planck correction (for a finite rotational temperature), as well as a correction for sub-thermal population in the upper rotational levels. While these corrections could vary from galaxy to galaxy, their scatter is unlikely to be greater than ±0.1 dex, as judged from recent data sets.

*2.1.4 Density-Temperature Dependence of the Conversion Factor.* This leaves the function $\zeta\left((<n(H_2)>)^{1/2}/T_{RJ}\right)$, which is correlated with the star formation rate at a given mass and redshift, that is, the vertical location in the stellar mass – star formation rate MS plane (Elbaz et al. 2011; Gracia-Carpio et al. 2011; Nordon et al. 2012; Lada et al. 2012). The average hydrogen gas density and temperature in MW GMCs may by $n\sim100$ and $T_{R1}\sim15$ K, such that $\zeta\sim O(1)$. Interestingly, in the much denser but also much warmer gas in extreme LIRG/ULIRGs in the local Universe, or in the gas rich SFGs at high-z, $\zeta$ may be broadly similar to that in the MW. From a comparison of the galaxy integrated data in the three different tracers discussed in below in section 3.2, this correction factor appears to be unity within ±0.15 dex within an offset $\delta MS=\log(SFR/(SFR(MS, z))$ of ±1.3 of the MS line at redshift z, $SFR(MS,z)$. With these uncertainties in the assumptions, calibrations and systematics the uncertainty of CO-based gas masses near the MS is likely no smaller than ±0.2 dex, and increasing away from the MS.

The cosmic microwave background, $T_{CMB} = 2.725\times(1+z)$, can become increasingly important for molecular excitations and radiative transport of molecular lines, the higher the redshift (section 2.10: Carilli & Walter 2013). First, the line emission is in the foreground of the CMB. Since the extended CMB continuum is resolved by the millimeter interferometers, the on-source line emission is lower than without the CMB, unless the line is optically thick. Likewise for optically thin gas the CMB increases the molecular excitation source function by $\delta S_{CMB} = (\exp(h\nu(J,z)/kT_{CMB})-1)^{-1}$. Since molecular gas densities and temperatures on galaxy integrated scales increase with redshift, the effects of the CMB overall are modest (5-20%) in terms of the source functions in the J=2 to 4 CO transitions and for z<3, and will be neglected in the following discussion. This may not be appropriate at high z.

## 2.2 The Dust Method: Far-IR SEDs

With the Herschel space observatory (2009-2013), several groups assembled deep far-IR continuum surveys with the PACS & SPIRE instruments (ATLAS: Eales et al. 2010; PEP: Lutz et al. 2011; GOODS-Herschel: Elbaz et al. 2011; HerMES: Oliver et al. 2012). Magdis et al. (2011, 2012b), Magnelli et al. (2014), Santini et al. (2014), Béthermin et al. (2015) and Berta et al. (2016) established 100 to 500µm far-IR SEDs from individual galaxies or from stacking PACS and SPIRE photometry in several of the cosmological deep fields (e.g. GOODS-N/S and COSMOS), for SFGs in the redshift range 0.1-2.5. As an example, Magnelli et al. (2014) binned their data onto a three dimensional grid in *z*, *SFR* and $M_*$, and stacked the photometry in each bin. They then fitted model SEDs from the library of Dale and Helou (2002), for which dust temperatures were established from single optically thin, modified blackbody fits with emission index β=1.5. From these stacked SEDs, they derived dust masses from Draine & Li (2007) and modified black body models (G15; Berta et al. 2016). Magdis et al. (2012b), Santini et al. (2014) and Béthermin et al. (2015) obtained similar results based on different galaxies and somewhat different methodologies. Berta et al. (2016) present a comprehensive analysis of uncertainties in Herschel-based dust masses. Comparing these different results the systematic uncertainties in dust masses obtained from this far-infrared technique is probably ±0.25 dex.

The conversion to gas masses requires a metallicity dependent dust-to-gas ratio correction, which also enters the redshift evolution through the redshift dependence of the mass-metallicity relation (e.g. Béthermin et al. 2015). Following Magdis et al. (2012b), Draine & Li (2007) model dust masses are converted to (molecular) gas masses by applying the Leroy et al. (2011) metallicity dependent gas to dust ratio fitting function for *z*~0 SFGs, $\delta_{gd} = M_{mo\lg as}/M_{dust} = 10^{(+2-0.85\times(12+\log(O/H)-8.67))}$ , where 12+log(O/H) again is the gas phase oxygen abundance (see also Draine et al 2007 for dust-to-gas with metallicity scalings of the SINGS nearby galaxy sample, and Galametz et al (2011) or Rémy-Ruyer et al. (2014) for lower metallicity galaxies down to 12+log(O/H)=8.0).

The gas column densities in the Leroy et al. (2011) recipe refer to the sum of molecular and atomic gas, but we are interested in the molecular gas ($H_2$) to dust ratio. From WISE/Herschel mid- and far-IR data and IRAM 30m CO 2-1 observations in 78 nearby SFGs in the 'stripe 82' region Bertemes et al. (2018) find a linear relation between CO-based (see above) and far-IR dust based columns and argue that for the molecular gas to dust ratio $\delta_{gd}$(molgas)~67, with little dependence on metallicity. In our analysis here, we use the gas-to-dust ratio analysis of Bertemes et al. (2018).

## 2.3 The Dust Method: 1mm Continuum Luminosity

Scoville et al. (2014, 2016) and S17 have proposed that a single frequency, broadband measurement in the Rayleigh-Jeans tail of the dust SED (for instance at 345 GHz, ~1mm) is sufficient to establish dust and gas masses. This Ansatz is justified if the emission is optically thin and the variation of the mass-weighted dust temperature on galactic scales is small. This is broadly consistent with the slow changes of average $T_{dust}$ with redshift and specific star formation rate near the MS in the stacked Herschel data (Magnelli et al. 2014), but larger and a wider range of temperatures are observed in high surface brightness galaxies, such as local ULIRGs (section 5). Based on SCUBA observations of a subset of the sample of $z\sim0$ disks from Draine et al. (2007), Scoville et al also argue that the molecular gas to dust ratio does not vary significantly with metallicity, in agreement with Bertemes et al. (2018). To calibrate the dust opacity at the observing frequency $v_{obs}$, Scoville et al. use a variety of CO observations of local normal and infrared luminous galaxies, as well as high-z submillimeter selected galaxies (SMGs), and the assumptions of $T_0$=25 K, $\alpha_0$=6.5 and $\delta_{gd}$=150 to determine the zero point $\alpha_{dust0}$(352 GHz)=$(L_{dust}/M_{molgas})|_{352\ GHz}$=$6.7\times10^{19}$ erg/s/Hz/$M_\odot$ (see Table 5 of Scoville et al. 2016 for references). This then yields

$$\left(\frac{M_{molgas}}{1\times10^{10}M_\odot}\right) = \left(\frac{S_{vobs}D_L^2}{mJy\times Gpc^2}\right) \times (1+z)^{-(3+\beta)} \times \left(\frac{v_{obs}}{352\ GHz}\right)^{-(2+\beta)} \times \left(\frac{6.7\times10^{19}}{\alpha_{dust0}}\right) \times \left(\frac{\delta_{gd}}{150}\right) \quad (3),$$

For the frequency dependence of the dust opacity Scoville et al. adopt $\beta$=1.8. Recently Kaasinen et al. (2019) have recalibrated this method through observations of both CO (1-0) and 850 micron dust emission in 12 $z\sim2$ SFGs, and find good agreement with the calibration of Equation (3).

In our analysis below (based on $\alpha_0$=4.36 and $\delta_{gd}$=67) we find best agreement between the available 1mm dust measurements and the far-IR and CO observations for $\alpha_{dust,0}$=$8\times10^{19}$, very close to the value proposed by Scoville et al. This is because, to first order, the larger CO-conversion factor of Scoville et al. compensates for their larger adopted gas-to-dust ratio.

## 2.4 Molecular Hydrogen and Other Molecular Lines

Molecular hydrogen, $H_2$, comprises most of the mass of GMCs. Yet because of its lack of an electric dipole moment, the rotational transition probabilities are very small, and the line emission is weak. Moreover, the low moment of inertia results in a wide spacing of the rotational states. The J=2-0 transition is at 28μm, equivalent to a level spacing of 500 K. As a result, the rotational transitions are only sufficiently excited in very warm star forming regions or shocked and UV irradiated gas (Parmar et al. 1991; Richter et al. 1995). In cases of bright background sources, $H_2$ ro-vibrational transitions at 2μm are detectable in absorption (Lacy et al. 1994, 2017). $H_2$ can also be detected in the far-UV Lyman- and Werner- electronic bands in absorption against background stars or AGN (e.g. Shull et al.

2000; Rachford et al. 2002, 2009; Tumlinson et al. 2002; Gillmon et al. 2006). However, the FUV observations can only sample the diffuse ISM or translucent clouds. In general, $H_2$ spectroscopy is not practical for surveying the cold gas content in galaxies.

While the lower CO rotational transitions predominantly probe moderately dense and cool gas, it can be of substantial interest to pick out the denser and more excited gas components in galaxies. For instance, the lower rotational transitions of HCN (Lee et al. 1990) can be bright in dense star forming clouds. Gao & Solomon (2004), Vanden Bout et al. (2004) and Gao et al. (2007) pioneered HCN observations in local luminous and ultra-luminous infrared galaxies (U)LIRGs and some very bright or lensed distant SFGs and SMGs. They found a linear relationship between HCN and far-IR luminosities, and 3-10 times enhanced HCN/CO flux ratios in these extreme starbursts as compared to normal spirals, and proposed that this excess is due to a highly elevated dense gas fraction. There are as yet no systematic studies of HCN (or other high-dipole moment molecules) at high-z, but these are now in principle feasible with the capabilities of ALMA and NOEMA.

Regions where CO is photo-dissociated by UV-photons, or destroyed by cosmic rays, might instead be studied in atomic or ionized carbon fine structure lines (Stacey et al. 2010; Tomassetti et al. 2014; Herrera-Camus et al. 2015; Bisbas et al. 2017; Papadopoulos et al. 2018).

## 2.5 Three Measurement Approaches

There are three approaches of studying the cosmic gas evolution:

**The 'pointed source' approach** starts with a flux/luminosity or mass selected parent sample, with well-established and homogeneously calibrated galaxy parameters (($M_*$, *SFR*, $R_e$). The SDSS sample is the prime example for $z$=0 (Brinchmann et al. 2004; Saintonge et al. 2011 a, b). At 0<$z$<3 the CANDELS/3D-HST surveys in the GOODS N/S, EGS, COSMOS and UDS fields are the currently most powerful such imaging samples (Grogin et al. 2011; Koekemoer et al. 2011; Brammer et al. 2012; Skelton et al. 2014; Momcheva et al. 2016). An appropriately selected sub-set of these parent samples then serves as a benchmark sample for the entire parent population. To the extent possible, CANDELS/3D-HST contains mid- (Spitzer/WISE) or far-IR (Herschel) photometry, such that star formation rates based on IR data are included (Wuyts et al. 2011a). The leading pointed surveys of the last decade, such as PHIBSS, have all emphasized massive galaxies, mainly to avoid large and uncertain corrections for sub-solar metallicities. Far-IR/submillimeter selected galaxy surveys (such as SHADES: Mortier et al. 2005; LESS: Weiss et al. 2009; Herschel-ATLAS: Eales et al. 2010) are generally much smaller, but pick out dusty massive SFGs, and SFGs above the MS. The hidden assumption of the pointed source approach is that the parent imaging survey contains essentially all relevant objects in the sky, so that only the quantity 'gas mass' must be established. The work of Rodighiero et al. (2011, 2015) shows that this assumption is valid to at least $z$~3.

**The 'deep field scanning'** approach usually selects a region on the sky, typically several square arc-minutes, for which multi-band data may be available, and blindly images the region for line or continuum emission, by stepping or scanning. Detections are then classified by reliability and matched to existing multi-band data (e.g. Decarli et al. 2014, 2016; Walter et al. 2016; Aravena et al. 2016, 2019; Dunlop et al. 2017; Pavesi et al. 2018; Riechers et al. 2019; Gonzalez-Lopez et al. 2019). This technique detects already known galaxies that happen to be in the survey field, and as such is equivalent to the pointed method. In addition, it is capable of detecting galaxies that were not present in the standard optical/UV/near-IR imagery, so is considered less 'biased' than the pointed method. As has been demonstrated by the highly productive HST deep fields (e.g. Williams et al. 1996; Giavalisco et al. 2004; Beckwith et al.2006), the deep field technique is preferred if the source density is high enough that source multiplicity per pointing (or per primary telescope beam for interferometers) is high.

The current efficacies of these two methods can be compared from published studies. The PHIBSS 1 and 2 CO surveys at NOEMA of AEGIS and CANDELS/3D-HST galaxies have average source detection times of 25 and 12 hours (>4$\sigma$, per detected galaxy, including calibration and overheads). The improvements are due to the increasing numbers of antennas and the quality of receivers between 2008 and 2018. Extrapolating to the NOEMA 12 array and to ALMA gives an average detection time for CO emission (>4$\sigma$) of *log ($M_*/M_\odot$)*=10.3-11.3 galaxies at *z*=0.7-2 of 4-6 hours (NOEMA 12) and 1.2-1.7 hours (ALMA 43), respectively, including calibration and overheads. For the ALMA-ASPECS *z*=1-2.6 deep field CO project of HUDF-S (DeCarli et al. 2019; Aravena et al. 2019) the average detection rate is 1 galaxy (**with a known optical counterpart**) per 4 hours of ALMA on source integration time. Most of the Aravena detections are at *z*~1-2.6 and *log ($M_*/M_\odot$)*=10-11.2, comparable to the assumptions above; there are two sources below *log ($M_*/M_\odot$)*=10. This detection rate is 2.7 times slower than ALMA pointed observations, probably caused by extra overheads in the deep-field approach, combined with the low surface density of massive SFGs (see below). The number of line **candidates** (without known counterparts) in all deep field projects is about twice as high as the number with counterparts, so the detection rate is 1 per 2 hours, assuming all these other detections are real. This is comparable to the ALMA pointed efficiency but then requires (HST, JWST…..) follow-up to establish the basic galaxy parameters. Given the cosmic volume covered by ASPECS 3mm, the number of detected sources (including less secure source candidates) is approximately what is expected from the MD14 mass functions at z=1-2.5 at high stellar masses. We conclude that the deep-field CO-detection efficiency (of ALMA) in the *log ($M_*/M_\odot$)*=10-11.2 bin is comparable to or slightly faster than NOEMA pointed observations, but about 2 to 3 times slower than pointed ALMA observations, assuming similar conditions.

The number of low mass galaxies (in the *log ($M_*/M_\odot$)*=9-10 bin) should be an order of magnitude greater than the few sources actually detected, indicating that the CO detections are substantially incomplete in that lower mass bin, and making the deep field CO technique much less efficient than it could be in principle.

The situation is comparable for 1mm dust observations. The detection rate at 1mm for pointed observations at ALMA is 1 galaxy per 10 minutes in the *log ($M_*/M_\odot$)*=10-11.2 bin and between $z$=1-2.6 (e.g. Scoville et al. 2016; Tadaki et al. 2017; Cowie et al. 2018). The Dunlop et al. (2017) 1mm deep field study detected one galaxy in these ranges per 1.5 hours, and only 3 galaxies below $10^{10}$ M$_\odot$, where the expectation would be ~46 from the MD14 mass functions. The dust technique is also incomplete at low masses.

**The third method is 'intensity mapping'**, the measurement of the wavelength power spectrum of the distant molecular medium without spatial imaging (Righi et al. 2008; Lidz et al. 2011; Mashian et al. 2015; Li et al. 2016; Breysse & Rahman 2017). This technique derives its heritage from similar efforts to detect the power spectrum of 21cm HI in the re-ionization epoch (PAPER: Ali et al. 2015; HERA: DeBoer et al. 2017). At the time of writing this review, no CO results have been reported from this technique.

All three techniques are challenged by the strong metallicity dependence of the CO conversion function $\chi(Z)$ below $Z \leq 0.5$ $Z_\odot$, due to CO photo-dissociation by FUV (e.g. Nordon & Sternberg 2016), and by the metallicity dependent gas to dust ratio. At $z$~2, a galaxy of mass $10^{10}$ M$_\odot$ has $Z$~0.4 $Z_\odot$ and two times fainter CO lines than a massive SFG, and a $10^9$ M$_\odot$ SFG has $Z$~0.13 $Z_\odot$ and 8 times fainter CO lines. This problem gets worse with redshift, such that it would be extremely difficult to detect CO emission in low mass $z$=2-8 SFGs. Metallicity effects on the dust continuum emission are more uncertain (see sections 2.2 and 2.3).

# 3. Galaxy Integrated Scaling Relations

## 3.1 Depletion Time as the Primary Parameter of Gas Evolution

Given the close connection of the molecular gas evolution to the cosmic star formation history and the star formation MS, a very important quantity is the ratio of the molecular gas mass to the *SFR*. This quantity is called the '**molecular depletion time scale**', $t_{depl} = M_{molgas}/SFR$ (Gyr), expressing the time in which a molecular gas reservoir would be depleted by the current star formation activity, not considering any mass return to the ISM from stellar winds and supernovae. The inverse of the depletion time scale, $\varphi_{depl}=1/t_{depl}$, is often described in the literature as an 'efficiency' (although it has units of 1/time), where $\varphi_{depl} = \varepsilon_{ff}/t_{ff}$ and $\varepsilon_{ff}$ is the efficiency of star formation per free-fall time or per galaxy dynamical time (see also section 3.5). The depletion time is the ratio of abscissa and ordinate in galaxy-integrated versions of the Kennicutt-Schmidt (KS) scaling relation between (molecular) gas and star formation rate (Kennicutt 1998; Kennicutt & Evans 2012). If $t_{depl}$ at a given $z$ does not vary with *SFR*, the slope of the KS relation, $N_{KS}$, is equal to unity (Bigiel et al. 2008; Leroy et al. 2008, 2013; Kennicutt & Evans 2012; Genzel et al. 2010; Daddi et al. 2010b). Once the dependence of the depletion time scale on redshift and the key galaxy parameters ($M_*$, *SFR*, $R_e$) are determined, the cosmic evolution of the gas to stellar mass ratio, $\mu_{molgas}=M_{molgas}/M_*$, follows automatically, since $\mu_{molgas}= t_{depl} \times sSFR$, and finally $f_{molgas}= \mu_{molgas}/(1+ \mu_{molgas})$.

G15, T18 and S17 have demonstrated that the dependencies of $t_{depl}$ and $\mu_{molgas}$ on the physical parameters can be separated as products of power laws that are functions of $(1+z)$, of offset from the MS, $\delta MS$ (perpendicular to the MS), and of $M_*$ (along the MS),

$$t_{depl} = t_0 \times (1+z)^A \times (SFR/SFR(MS,z,M_*))^B \times (M_*/5\times 10^{10} M_{sun})^C \quad (4).$$

As we show below, studies of cosmic cold gas evolution agree that the galaxy integrated depletion time scale depends mainly on redshift and offset from the MS (at a given $z$). There is little or no dependence on galaxy mass, size or environment. Since >90% of the cosmic star formation rate occurs on the MS (Rodighiero et al. 2011, 2015) **the single-parameter sequence $t_{depl}(z)|_{MS}$ can be used as a first order estimate to determine the cosmic evolution of cold gas, if the star formation history is well known**. We feel that this is a very important conclusion, since the measurement of a molecular gas mass of a galaxy is much more costly than determining a star formation rate. Currently there are a few thousand published galaxy gas mass measurements, compared to $10^6$ or more star formation rates.

## 3.2 Depletion Time Scaling Relations

To update the depletion time scaling relations of G15, T18 and S17, we have assembled from the existing literature and the ALMA archive molecular gas mass detections (>3.8$\sigma$) for individual galaxies and stacks (see section 2.2 and Table 1) for 2052 SFGs between $z=0$ and $z=5.3$, $logM_*=9$ and 12.2, and $\delta MS=-2.6$ and +2.2 (assuming the S14 definition of the MS). $SFR$s range between $logSFR=-1.5$ and +3.75. Of the entries in this compilation[4], 858 are based on CO (1-0, 2-1, 3-2, or 4-3) detections (see section 2.1 for details), 724 on far-IR dust measurements (section 2.2) and 470 on ~1mm dust measurements (section 2.3). We list the surveys and references for this compilation in Table 1. The dominant uncertainties are systematic, rather than statistical uncertainties of individual measurements, which are typically <±0.1 dex. Assumptions on IMFs and star formation histories dominate for stellar masses and star formation rates and result in ±0.15 to ±0.25 dex uncertainties, increasing to ±0.3 dex above the MS, in the starburst regime. The various assumptions going into the derivation of CO- or dust-based molecular masses are ±0.25 dex. Some of these systematic dependencies drop out when using reduced quantities, such as $sSFR$, or $\mu_{molgas}$, since numerator and denominator are affected similarly.

---

[4] The full list will be publicly available and accessible upon publication of the article. The list contains the basic ancillary information ($z$, $logM_*$, $log\,SFR$, $log\,sSFR$, $log\,t_{depl}$, $log\mu_{molgas}$, $R_e$(*rest frame optical*)) and the zero point offsets chosen for the input $log\,t_{depl}$ and $log\mu_{molgas}$. We request the reader to refer to this article when using these data, or to the original papers, which are also listed.

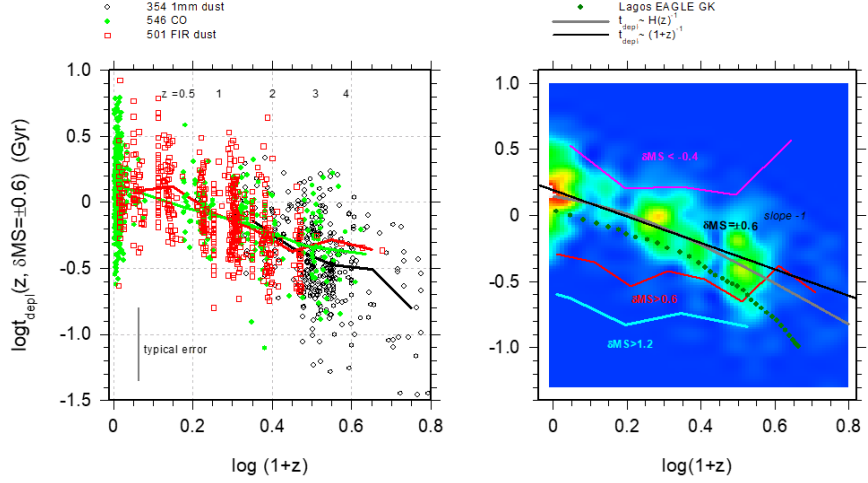

*Figure 3. Molecular depletion time scale (Gyr) as a function of redshift, for 1401 MS SFGs ($\delta MS = \pm 0.6$) in the master sample. Left: CO(green), far-IR dust (red) and 1mm dust (black) measurements after application of the small zero point corrections to each data set (between -0.1 and +0.27 dex), in order to minimize the overall scatter. The green, red and black lines are the data averages within bins of about 0.15 dex. The scatter of the residuals around the best fit, slope -1 power law is $\pm 0.21$, $\pm 0.18$ and $\pm 0.26$ dex for the CO, far-IR dust and 1mm dust measurements, respectively. For CO and far-IR data, the scatter is well within the combined statistical and systematic errors. The larger 1mm residuals may in part arise from the fact that many galaxies only have photometric input redshifts, which can be uncertain at z>2. Judging from the trend lines overall the three different techniques are in good to very good agreement, but at z>2 the 1mm dust data are ~0.15-0.2 dex below either of the far-IR or CO data sets. This is reflected in slightly different inferred slopes $B=dlogt_{depl}/dlog(1+z)$ if the data sets are considered separately. CO and far-IR data yield B=-0.88 ($\pm 0.1$), -1.1 ($\pm 0.16$) while the 1mm data give B=-1.3 ($\pm 0.32$). This barely significant tension is also seen in the analysis of T18, Table 3 and S17, Table 2. The quoted uncertainties are 2 $\sigma$ fit errors. The combined data set has B=-0.98 ($\pm 0.06$). The zero point of the relation is $logt_0=0.21$ ($\pm 0.03$), or $t_0=1.6$ ($\pm 0.5$) Gyr. Right panel: The best fit for the combined MS-data (black, slope -1) is superposed on the smoothed overall distribution function including all three MS data sets shown in color, and compared to the above/below MS outlier trend-lines: $\delta MS > +0.6$ (red), $\delta MS > +1.2$ (cyan), $\delta MS < -0.4$ (magenta). These trend lines capture the $(SFR/SFR(MS))^{-0.5}$ dependence of the depletion time scale. The thicker continuous grey line shows that an equally good fit to the MS-data set can be obtained with an Ansatz $t_{depl} \sim H(z)^{-1} \sim 1.6$ Gyrs $\times (0.7 + 0.3 \times (1+z)^3)^{-1/2}$. Finally, the thick dotted green line is the redshift dependence of the depletion time obtained by Lagos et al. (2015) from their post-processing analysis of the EAGLE simulation. The overall shape is consistent with the observations but the zero-point ~0.15-0.3 dex too low. The dashed black line is the best fit of S17 (B=-1.05 ($\pm 0.1$)).*

### 3.2.1 Redshift Scaling.
Following T18, we first establish zero-point correction factors for each set of measurements, relative to the final average. We consider only the MS-SFGs at all redshifts ($\delta MS=\pm 0.6$) and exploit the finding that this subset varies only as a function of redshift. We then establish zero-point offsets for the various data sets (between -0.1 and

+0.27 dex) to minimize the scatter of the *log $t_{depl}$-log(1+z)* relation. Figure 3 (left) shows all the resulting MS points individually and in different colors for the three gas mass determination techniques. For all three methods the redshift trend for MS galaxies can be well fitted by a power law of slope B=*d log $t_{depl}$/d log(1+z)*= -0.9 (±0.1, CO), -1.1 (±0.12, far-IR) and -1.3 (±0.32, 1mm), where all uncertainties quoted here and below are 2σ fit errors. The combined slope for all 3 techniques is B= -0.98 (±0.06, ±0.1), where the first is the statistical uncertainty, and the second is the systematic error, depending on the weighting of different data sets, or whether all data points are used, and simultaneously fitting t0, A, B, C and D. This value for the redshift slope is in excellent agreement with, and improves on T18, S17, G15, Daddi et al. (2010b), Magdis et al. (2012b), Sargent et al. (2014), Santini et al. (2014) and Béthermin et al. (2015).

*3.2.2 Scaling perpendicular to the MS.* Next, we consider the variation of depletion time perpendicular to the MS line. Because of the separation Ansatz of G15, S17 and T18, it is possible to analyze the entire 2052 data points simultaneously, after removing the redshift dependence. This greatly increases the statistical robustness and the range in *δMS*. In the right panel of Figure 3, we show the binned averages of $t_{depl}$ *(z)* in two bins above and one bin below the MS. There is a clear and continuous trend at all redshifts for the depletion time to decrease above, and increase below the MS. The formal fit yields *C=dlog$t_{depl}$/δMS*=-0.49(±0.03). This value agrees very well with Saintonge et al. (2011b), Huang & Kauffmann (2015), G15, Scoville et al. 2016 (*C*=-0.55±0.1) and T18 but there is some tension with S17 who find a steeper dependence (*C*=-0.7 (±0.04)), or with Magdis et al. (2012b), Sargent et al. (2014) & Santini et al. (2014) who find a shallower dependence (*C*=-0.1..-0.4).

**A very important general conclusion is that all three techniques, (a) CO line luminosity, (b) far-IR dust SED, and (c) 1mm dust photometry, yield remarkably similar scaling relations, once the zero points are dealt with through cross-calibration** (Figure 3, 4, 5; G15; S17; T18; Magdis et al. 2017). This enables searching for systematic trends possibly affecting molecular mass measurements in any individual one of these three methods. The slight tension between the CO-based and the dust-based estimates of C (shallower slope for the CO-based data) might suggest that there is a redshift-dependence of *$α_{CO}$*. To alleviate this tension to first order, we average the estimates of the three techniques. Note that our method of averaging will fail if their zero points co-vary in similar fashion. When considering the quoted final parameter uncertainties for a given data set it is important to keep in mind the coverage in the specific parameter. T18 have shown from mock data sets that limited redshift-coverage, in addition to small number statistics, can strongly limit the final precision of parameter estimates. Covariance between parameters also needs to be taken into account. As the result of these co-variances and the overall calibration uncertainties, zero points of all individual techniques are no better than ±0.2 dex.

From these crosschecks, we deduce that the slope *C* does not depend significantly on redshift, nor on *δMS*. This is shown in Figure 4. We note that we have deliberately removed a slope of -0.5×*δMS* in the two panels of Figure 4, instead of the best fitting slope of -0.49×*δMS*. This difference can be easily seen by eye as a significant positive slope of the residuals near the MS. Figure 4 demonstrates that the residuals stay fairly flat across the entire range of $\delta MS$ sampled, from below the MS $(\delta MS \sim -1)$ to the extreme starbursts above the MS $(\delta MS \sim +2)$.

Based on the comparison of dynamical and gas masses in ULIRGs Scoville et al. (1997) and Downes & Solomon (1998) concluded that there is a significant drop of the CO conversion factor $\alpha_{CO}$ from the MS to extreme starburst galaxies ($\alpha_{CO}$=0.8…3) This drop in $\alpha_{CO}$ has been a standard assumption in the field since then, including in our own work (e.g. Genzel et al. 2010) and in Bolatto et al. (2013). In G15, T18 and here **the low gas masses inferred from the gas dynamics in ULIRGs, and more generally above the MS, are now encapsulated in the $\delta MS^{-1/2}$ dependence of $t_{depl}$, instead of a change in $\alpha_{CO}$.** No additional variations of $\alpha_{CO}$ with $\delta MS$ are needed; S16 and S17 have come to the same conclusion. The overall 50% larger gas masses proposed by S16 and S17 ($\alpha_{CO}$= 6.5 instead of 4.36) are within the systematic uncertainties of the CO zero points, but our lower values are in better agreement with the dust techniques and recent simulations (section 4.2.1). Figure 4 also suggests that any underlying changes in the physics of star formation as one moves from the MS-line upward to the starburst populations, or down to the passive population, are most likely gradual, and not bi-modal, as advocated by Genzel et al. (2010), Magdis et al. (2012b) and Sargent et al. (2014).

This also means that below the MS, the slope of the $t_{depl}$ (MS) does not change significantly (dotted lines in the left panel of Figure 4 show the uncertainties per $\delta MS$ bin). This conclusion is still uncertain because of the small number of galaxies below the MS (individual data points in the right panel of Figure 4), and the larger scatter of the 1mm data (black circles), but could be important input into the question of whether different physical processes, such as morphological quenching, are important below the MS (Genzel et al. 2014, Suess et al. 2017, Spilker et al. 2018).

The negative slope of the $t_{depl}$-$\delta MS$ relation (C~-0.5) affects the slope of the KS-relation in samples of galaxies that contain a mixture of MS-galaxies and starbursts (see section 4.4 in G15). Since the KS-ordinate (*SFR*) scales inversely with $t_{depl}$, starburst galaxies ($\delta MS \sim +1…+2$) at a given molecular gas mass (surface density) lie above the MS-galaxies, thus steepening the slope of the KS-relation above unity (for constant $t_{depl}$). This explains plausibly why Kennicutt (1998) and Kennicutt & Evans (2012, Figure 11) find $N_{KS}$=1.4 and Daddi et al. (2010b) find $N_{KS}$=1.3-1.4 at high $\Sigma_{gas}$, while Bigiel et al. (2008) and Genzel et al. (2010) find $N_{KS}$= 0.9-1.2. The former papers include substantial numbers of starbursts, ULIRGs and SMGs, while the latter papers focus on MS-galaxies. These results are all consistent once the dependence of $t_{depl}$ on $\delta MS$ is considered (G15).

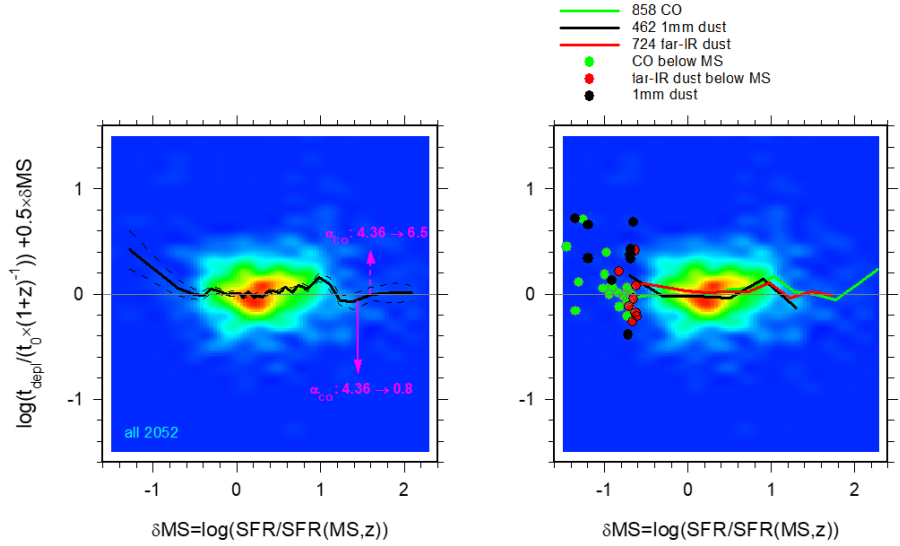

***Figure 4.*** *Residuals of depletion time as a function of MS offset, after removing both the redshift dependence, as well as a slope $C=dlogt_{depl}/\delta MS=-0.5$ (note the best fit slope is -0.49 ($\pm 0.03$)). The color distribution contains all 2052 data points in our master list. Left: The black line is the binned median of these data, and the dashed lines depict the $1\sigma$ statistical errors. The 'over-rotation' of the original data by applying a slope of -0.5 is quite visible in the residuals. The down pointing magenta arrow shows the change of the CO conversion factor would change from $\alpha_{CO}=4.36$ at the MS to 0.8 at $\delta MS=1.5-2.2$, as proposed by Downes & Solomon (1998). The lower masses above the MS and in ULIRGs implied by that change in $\alpha_{CO}$ is captured in our scaling relations by the $(SFR/SFR(MS))^{-0.5}$ dependence of $t_{depl}$ instead. The up-pointing dotted arrow denotes $\alpha_{CO}=6.5$ instead of 4.36, as proposed by Scoville (2014, 2015, 2016), S17. Right: Green, red and black continuous lines mark the binned averages of CO, far-IR dust and 1mm dust. These findings (see also G15, T18, S17) call into question the possible existence of a sharp transition from MS galaxies to starbursts (as a result of variations in $\alpha_{CO}$) and the existence of a bi-modal star formation distribution, as advanced by Genzel et al. (2010), Daddi et al. (2010b), Magdis et al. (2012b) and Sargent et al. (2014). Likewise, any change across $\delta MS=log(SFR/SFR(MS))$ of the CO conversion factor should be apparent as a slope difference between the three different tracers. Such differences are not present, excluding a change in $\alpha_{CO}$ between MS and starburst galaxies by more than 0.15 dex. Green, red and black circles denote CO, far-IR and 1mm observed galaxies below the MS.*

*3.2.3 Dependence on other Parameters.* Finally, considering the $t_{depl}$ residuals after removing both the *z* and *δMS* dependencies, or carrying out a global overall fit, the mass and radius dependence is negligible with $D=dlog\ t_{depl}/dlogM_*=0.03\pm0.04$ and $E=dlog\ t_{depl}/dlogR=0.09\pm0.15$, in good agreement with the publications cited in the last paragraph. Table 2a summarizes these best fitting functions.

## 3.3 Scaling Relations for Molecular to Stellar Mass Ratios

Having established the parameter dependences of $t_{depl}(z,\delta MS,M_*,R)$ we can multiply the depletion time scaling relations by Equation (1) and obtain $\mu_{molgas}(z,\delta MS,M_*) = M_{molgas}/M_*$. Alternatively, one can start with the individual molecular masses and stellar masses and then proceed by local or global fitting, with the same separation Ansatz as above for the depletion time. Figure 5 depicts the dependencies of $\mu_{molgas}$ on these three variables in graphical form, as well as the z-dependence of the molecular gas fraction, $f_{molgas}=\mu_{molgas}/(1+\mu_{molgas})$.

Molecular mass to stellar mass ratios and molecular gas fractions broadly track the redshift dependence of specific star formation rates (Equation 1), since the steep $(1+z)^3$ dependence of *sSFR* dominates over the slower redshift dependence of $t_{depl}$. As for *sSFR(z)*, the redshift dependence of $\mu_{molgas}$ is better described by a quadratic function (see T18). MS galaxies had the largest molecular gas fractions at $z\sim2-3$. Since $\mu_{molgas} = t_{depl} \times sSFR$ the slope=-0.5 anti-correlation of $t_{depl}$ with *δMS* is mirrored into a positive, slope +0.5 correlation of $\mu_{molgas}$ with *δMS*. The drop in *sSFR* with stellar mass is fully reflected in the relation between $\mu_{molgas}$ and stellar mass. The values for the best fitting function are given in Table 2b.

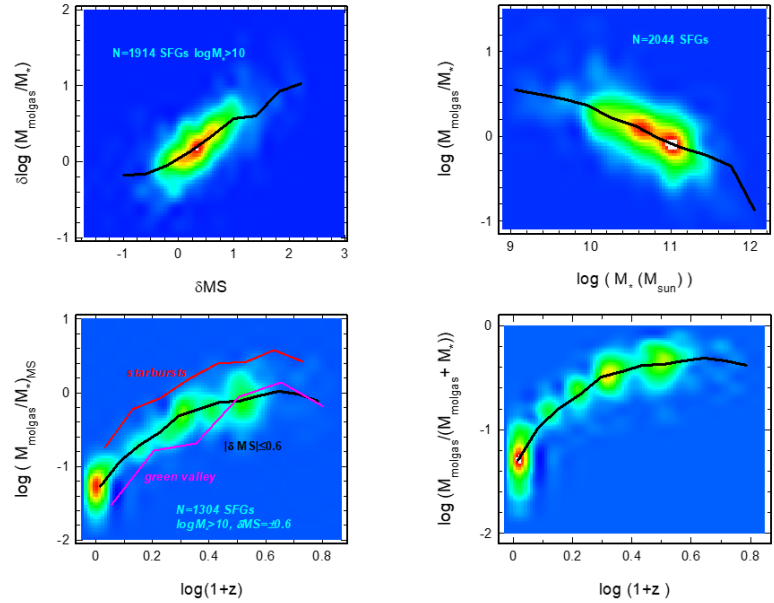

***Figure 5.*** *Scaling relations of $\mu_{gas}=M_{molgas}/M_*$ with redshift (bottom left), specific star formation rate offset δMS (top left) and stellar mass (top right), and molecular to total mass fraction $f_{molgas}=\mu_{molgas}/(1+\mu_{molgas})$ (bottom right). The smoothed distributions of the galaxies in the master set are shown underlying in color. In all panels the thick black line marks the binned averages of the overall distribution in bins of 0.1..0.25 dex. In the bottom left panel the black line and underlying color distribution refer to the MS (δMS=±0.6). The red and magenta lines denote the corresponding averages for starbursts (δMS>0.6) and below MS galaxies (δMS<-0.4).*

## 3.4 The Role of Environment

The scaling relations derived above refer to field galaxies, and not members of dense, massive clusters. For a complete picture of galaxy gas properties, we must also investigate whether environment is an additional parameter regulating the scaling of the gas and star formation properties. To date two approaches have been used to study this effect: the first is to calculate the over-densities of galaxies in large molecular gas and dust surveys; the second is to observe galaxies located in massive clusters or proto-clusters, and then compare their gas masses, gas fractions and depletion times with those of field galaxy samples. Darvish et al. (2018) have used the first method to analyze the S17 sample and a local HI-based sample from the ALFALFA survey (Haynes et al. 2011) to determine the scaling of $f_{molgas}$ and $t_{depl}$ with galaxy over-density. They separate the samples into 4 density bins, and match z, $\delta MS$ and $M_*$ in each bin to those of the >100 galaxies in the highest density bin to get a fair comparison. They find no correlation of $f_{molgas}$ and $t_{depl}$ with galaxy over-density, and conclude that the molecular gas content and star formation activity of a galaxy is regulated by internal processes from $z$=0-3.5, and that galaxy environment plays little or no role. A caveat is that their study does not include galaxies that are located in the densest environments, with over-densities relative to the field of ~100 (e.g. Tadaki et al. 2019).

Pointed studies of cluster and proto-cluster galaxies from $z$=0 to $z$~2.5 have come to different conclusions on the role of environment, but are presently quite limited in sample sizes and possibly affected by selection effects. At $z$=0, Kenney & Young (1989) and Koyama et al. (2017) see no effect on the molecular gas content in cluster galaxies. However, Mok et al. (2016) find an excess of gas in galaxies in dense environments, and Boselli & Gavazzi (2014) find evidence for ram pressure stripping resulting in lower gas mass and gas fractions in dense cluster galaxies relative to field galaxies. At high redshift, studies are emerging that compare molecular gas contents of cluster/proto-cluster members with the scaling relations of T18, S17 and G15. These studies mostly find either consistent or enhanced gas fractions relative to these "field" scaling relations (Noble et al. 2017; Rudnick et a. 2017; Lee et al. 2017; Hayashi et al. 2018; Tadaki et al. 2019), and some also find longer depletion times, relative to T18, S17 or G15 (e.g. Hayashi et al. 2018). Tadaki et al. (2019) also find that there is a mass dependent environment effect in their study of 66 SFGs in three $z$~2.5 proto-clusters. Galaxies with $10.5 < log(M_*/M_\odot) < 11$ have higher gas fractions and longer depletion times relative to the T18 relations, but this effect vanishes at $log(M_*/M_\odot) > 11$. They postulate that gas accretion could be accelerated in less massive galaxies, and suppressed in the most massive halos, possibly due to inefficient cooling in $log(M_{DM}/M_\odot)>12$ halos (Rees & Ostriker 1977; Dekel & Birnboim 2006). Conclusive progress in assessing whether galaxy environment is another parameter in the scaling relations described in sections 3.2 and 3.3 awaits statistically significant samples of several thousand galaxies or more, spanning a large dynamic range in galaxy over-densities.

## 3.5 What sets the Depletion Time?

The right panel of Figure 3 shows that the data are also well fit by the functional form $t_{depl} = (\varepsilon_H \times H(z))^{-1} = \varepsilon_H^{-1} \times t_H \times (0.7 + 0.3 \times (1+z)^3)^{-1/2}$. Here $H(z)^{-1}$ is the Hubble time at $z$, and $t_H$=13.98 Gyr is the current Hubble time (grey curve in the right panel of Figure 3). In that case $\varepsilon_H^{-1}$=0.1 (±0.02).

Following Mo et al. (1998, henceforth MMW) we consider a rotationally supported baryonic disk inside a virialized dark matter halo at redshift $z$, where baryons and dark matter have comparable specific angular momenta $j_{baryon} \sim j_{DM}$, and where the angular momentum parameter of the dark matter halo is $\lambda_a$. Observations of the angular momenta of low- and high-z SFGs by Romanowsky & Fall (2012), Fall & Romanowsky (2013), Burkert et al. (2016) and Swinbank et al. (2017) suggest that $j_{baryon}/j_{DM} \sim 1$ and $\lambda_a \sim 0.037$. The dynamical time of the centrifugally supported baryonic disk at the half-mass radius $R_e$ then is

$$t_{dyn}(R_e) = R_e / v_c(R_e) = 6.15 \times 10^7 \left(\frac{j_{baryon}}{j_{DM}}\right) \times \left(\frac{\lambda_a}{0.037}\right) \times f_h \times f_{ac} \times (H(z)/H_0)^{-1} \quad (5),$$

where the factor $f_h = v_c(R_{virial})/v_c(R_e)$ is unity for an isothermal rotation curve and ~1.5-2 for an NFW distribution, and $f_{ac} \leq 1$ denotes whether the halo has experienced adiabatic contraction. We adopt $f_h \times f_{ac} \sim 1.5$. Next we assume that the depletion time scale is proportional to the Toomre time of the disk, that is, the fragmentation time of the largest unstable mode, which is also the vertical oscillation time, $t_{depl}=t_T/\varepsilon_T=t_{dyn} \times Q/\varepsilon_T$ (See sidebar, Behrendt et al. 2015, Burkert et al. 2019). We reformulate $Q$ described in the sidebar as $Q=1.41 \times (\sigma_0/v_c)/f_{molgas}$. For $\mu_{molgas}<1$, appropriate for the MS at $z<2$, $f_{molgas} \sim \mu_{molgas}$ to within 30-50% (Figure 5). Putting this all together, we can express the depletion time as

$$t_{depl} = 1.3 \times 10^8 \times \frac{\lambda_{0.037}}{\varepsilon_T} \times \left(\frac{j_{baryon}}{j_{DM}}\right) \times \left(\frac{f_h f_{ac}}{1.5}\right) \times \left(\frac{H(z)}{H_0}\right)^{-1} \times \left(\left\{v_c/\sigma_0\right\} \times \mu_{molgas}\right)^{-1} \quad (yr) \quad (6).$$

The redshift dependence of the depletion time scale (or the slope of the KS-relation on the MS) in this model is entirely in $H(z)$, in agreement with the discussion above and Figure 3, since $(v_c/\sigma_0) \times \mu_{molgas}$ is approximately constant and near-unity as a function of redshift (Wisnioski et al. 2015; Übler et al. 2019). At a given redshift and along the MS, the depletion time should not change much, since the slow drop of gas fractions with increasing stellar mass compensates for an equivalent shallow increase of $v_c/\sigma_0$ with mass. Finally, the dependence on $\delta MS$ is entirely in $\mu_{molgas} \sim \delta MS^{0.5}$ (Table 2), and $t_{depl} \sim \delta MS^{-0.5}$, as observed. We note, however, that the implication of the above discussion that the average $Q$ is lower above the MS and at high z, is strictly not consistent with the model of galaxies self-regulating to $Q \sim Q_{crit}$.

This simple model gives the intriguing result that **the gas consumption time is directly tied to the overall galactic 'clock', as would be expected in a marginally stable, Toomre $Q<0.67$ disk** (Elmegreen 1997; Silk 1997; Genzel et al. 2011; Krumholz et al.

2012). The galactic clock is tied to cosmic time, $H(z)^{-1}$, which drives the average size and dynamical time scale of the disk (MMW). Equation (6) assumes a constant angular momentum parameter $\lambda$. In reality, angular momentum parameters in galactic disks follow a log-normal distribution with a scatter in log $\lambda$ of about 0.2 (Burkert et al. 2016), such that equation (6) is valid only for broad averages. Since gas fractions increase upward in the $M_*$-SFR plane, galaxies above the MS have smaller $Q$ and thus smaller depletion time scales. We conclude that the simple model of a Toomre-instability controlled disk predicts the correct scaling relations, at least at high redshift. It does not predict the 'zero point' $\varepsilon_T \sim 0.1$, or the star formation efficiency per dynamical time ($\varepsilon_{ff}$), which are likely set on cloud scales (Krumholz et al. 2005; McKee & Ostriker 2007). The Toomre and global disk instability model (Genzel et al. 2008; Krumholz et al. 2012; Dekel & Krumholz 2013) breaks down in the local Universe, where the average Toomre parameter is significantly above unity ($Q_{gas} \sim 2$-8; Leroy et al. 2008; left panel of Figure 6). In this regime, the velocity dispersion is set by feedback rather than gravitational instabilities (Ostriker & Shetty 2011; Krumholz et al. 2018; Übler et al. 2019).

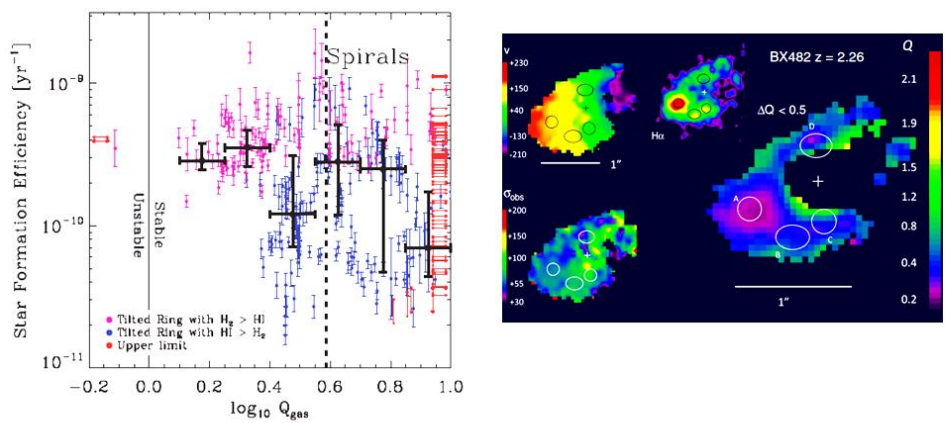

*Figure 6. Variations of the Toomre Q-parameter in galactic disks galaxies. Left panel: Star formation 'efficiency' ($1/t_{depl}$) as a function of $Q_{gas}$ in the z~0 galaxies of the HERACLES survey (adapted from Leroy et al. 2008). There is no significant dependence of $t_{depl}$ on Q, and all galaxies are in the 'stable' regime of the Q-parameter, such that these local star forming galaxies are likely not controlled by the Toomre global disk instability model. Right panel: Inferred $Q_{tot}$ parameter from the Hα dynamics in the z~2.26 disk/ring galaxy BX 482. Molecular column densities are inferred from the scaling relations (or inverted KS-relation). The Toomre Q parameter appears to be <1 everywhere in the disk, and the prominent star formation clump in the east represents a minimum in Q, suggesting that the global instability model is applicable (adapted from Genzel et al. 2011).*

The question remains whether the Toomre model is also applicable on sub-galactic scales. Leroy et al. (2013) have shown that the depletion time does not vary with dynamical time or radius in 30 local SFGs (left panel of Figure 6). This lack of correlation is consistent with the fact that $Q>1$ in these systems (Leroy et al. 2008). At high $z$, Genzel et al. (2011)

show that *Q*<1 holds everywhere in several massive star forming disks on >2 kpc scales, with clear minima at prominent massive star forming clumps (right panel of Figure 6). This suggests that the Toomre model may also be applicable in spatially resolved data of high-*z* SFGs but extensive, high resolution mapping at sub-clump scales will be necessary to for a definitive conclusion.

# 4. The Cosmic Evolution of Cold Gas Reservoirs

## 4.1 Mass Integrated Evolution

In the last section we discussed the steep increase with *z* of molecular gas reservoirs in SFGs, derived from pointed observations of massive ($log(M_*/M_\odot) >10.0$) systems. The next question is how the mass density of molecular gas per co-moving cosmic volume evolves with redshift or time. There are two paths to estimate this quantity. One is to use the "blind" technique of summing up the inferred molecular masses for a given redshift bin in the search volume (field area times times depth in units of co-moving Mpc). This blind approach has been taken by several authors in GOODS-N (DeCarli et al. 2014, *z*=1.5, 2.5) with NOEMA, HUDF (Walter et al. 2016, DeCarli et al. 2016: ASPECS, *z*=1, 2.3) with ALMA and in GOODS-N and COSMOS (Riechers et al. 2019: COLDz, *z*=2.5, 6) with the JVLA. Most recently, Lenkic et al. (2019) have analyzed the 110 data cubes of the NOEMA PHIBSS2 pointed CO survey (Freundlich et al. 2019; T18) of *z*~0.5-2.5 MS SFGs to search for additional serendipitous sources in the NOEMA primary beam field of view. They find 67 candidate sources, ~64% of which have potential optical counterparts. From these equivalently 'blind' data they determine the cosmic molecular mass volume density over redshifts from *z*~0.7 and 5. Figure 7 shows the compiled results of all these various studies. The molecular mass content per volume increases with increasing redshift, reaches a broad peak at $z_{peak}(molgas)$ ~1.4±0.3, and slowly drops toward higher redshift.

A second approach is to use the evolution of the SFR density (MD14), and multiply by the mass independent depletion time vs redshift scaling of the MS, $t_{depl}(MS)=1.6\times(1+z)^{-1}$, obtained from the pointed technique and discussed earlier (see Figure 3). This is justified since ~90% of the total cosmic SFR to *z*~2 occurs in MS galaxies (Rodighiero et al. 2011, 2015); outliers in the starburst population above the MS, with 3-10 times shorter depletion time scales, play only a secondary role in the overall cosmic evolution of the star formation rate and cold gas mass. The result is the thick blue curve in Figure 7. To within the uncertainties, driven by source number statistics in the deep field technique and the assumption of a single parameter dependence of the depletion time in the pointed technique, **the mass-integrated evolutions of the cosmic molecular gas density obtained with the two approaches are in very good agreement.** The peak redshift of galaxy molecular gas reservoirs is at somewhat lower *z* than that of the *SFR* density itself ($z_{peak}(SFR)$~2).

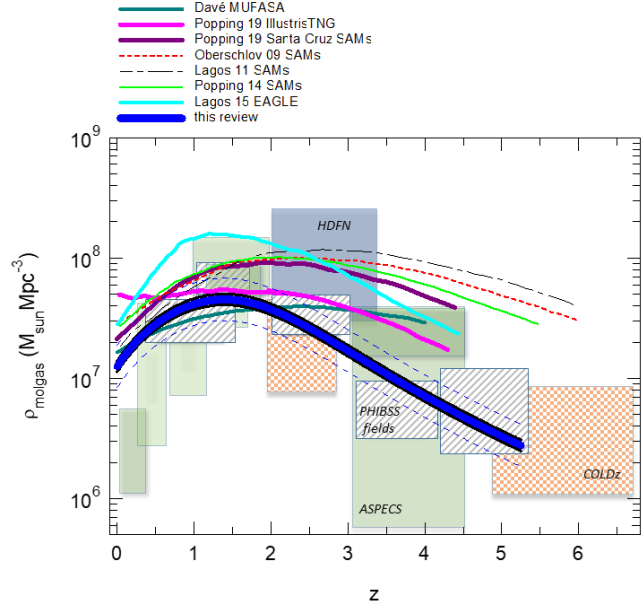

*Figure 7. Different observational and theoretical estimates of the cosmic evolution of total molecular mass (H$_2$ plus He) density per co-moving volume. The different boxes denote estimates at different redshifts with the 'deep field' technique (DeCarli et al. 2014, 2016, Walter et al. 2016, Aravena et al. 2016, Riechers et al. 2019). The grey-shaded boxes show the serendipitous detections of secondary sources in the same data cubes as those from the pointed PHIBSS2@NOEMA survey, which is another way to construct a 'blind' survey (Lenkic et al. 2019). The PHIBSS2 survey covers a larger cosmic volume than the other blind surveys, yielding smaller uncertainties. The thick blue-black line (with the parallel dotted blue lines denoting the uncertainties) is the star formation rate volume density of MD14 multiplied by $t_{depl}=1.6 \times (1+z)^{-1}$, which describes the consumption time of MS galaxies that dominate the star formation rate (Rodighiero et al. 2011, 2015). For comparison, the various lines denote the results of the SAMs and hydro-simulations in the recent literature.*

Essentially all semi-analytic models shown in Figure 7 predict that the maximum of the cosmic molecular gas density is at higher redshift, $z$~2-3. The post-processing analyses of the hydro simulations fare better. In particular, the two most recent simulation based results, from MUFASA (Davé et al. 2017) and IllustrisTNG (Popping et al. 2019) come close to predicting the observed time evolution and the maximum density. The right panel of Figure 1 shows the same curve from the depletion time scaling relation, now in comparison to the evolution of stellar mass from MD14. Filled red circles are individual measurements, and the thin red curve is the integration of the SFR in the left panel over time, for a Chabrier IMF and 40% mass return). Both Figures 1 and 7 show that the molecular gas reservoirs and stellar mass track each other from high-z to $z$~2, and then the molecular gas mass density levels out, soon thereafter followed by that of the stars.

We also show the average mass density of atomic hydrogen as a comparison. The neutral gas is probed by the 21 cm line in the local Universe and Lyα UV absorption spectroscopy in the mid- and high-z Universe (e.g. Keres et al. 2003; Zwaan et al. 2005; Peroux et al. 2005; Rao et al. 2006; Guimaraes et al. 2009; Obreschkow & Rawlings 2009a, b; Prochaska & Wolfe 2009; Martin et al. 2010; Braun 2012; Noterdaeme et al. 2012; Popping et al. 2014; Catinella et al. 2018; Rhee et al. 2018). Although no continuous "deep field" studies are yet available for HI, the atomic hydrogen distribution is quite flat ($\rho_{HI} \sim 5.6 \times 10^{+7} \times (1+z)^{0.5}$ Mpc$^{-3}$) and is decoupled from the rapid evolution of molecular or stellar mass and *SFR*. One possible concern is that the local galaxy HI measurements measure the gas within galaxies, while the absorption line measurements at higher redshifts probe gas in the galaxies, as well as in the CGM. We refer the reader to the article by Peroux & Howk in this volume for more details.

## 4.2 Molecular Gas Mass and SFR Density

The next topics are the evolution of mass and SFR distribution functions. Figure 8 summarizes the current observational and theoretical knowledge in molecular mass (left) and SFR (right) as a function of gas mass, in two redshift slices, *z*=0 (bottom) and *z*=2 (top). The left panels show molecular mass functions (per logarithmic mass interval) from the "metallicity" based recipes of Popping et al. (2014), based on the Gnedin (2000) and Kravtsov et al. (2004) (GK) treatment of re-ionization, baryonic collapse and $H_2$ formation. The Popping et al. (2014) and Lagos et al. (2015) models predict a broad mass function at all three redshifts, with a peak at ~7-10×10$^9$ M$_\odot$, and a broad shoulder toward lower masses, so that the median mass is less than half the peak mass. We compare these predictions to the observed *z*=2.2 mass functions, from a combination of ASPECs (filled green recatangles) (DeCarli et al. 2016), COLDz (Riechers et al. 2019), and PHIBSS2 deep field data at that redshift (adapted from Figure 13 of Lenkic et al. 2019), Herschel far-IR continuum data (black crossed rectangles, Vallini et al. 2016), and conversion of the MD14 SFR data to molecular masses with the depletion time scalings (filled blue circles). While data and models agree reasonably well at z=0 possibly in part because of the calibration of SAM parameters on the *z*=0 data.

### *4.2.1 Tension between Observations and Models at High Masses.*

At *z*=1-2 there is significant tension between the observed and theoretically predicted molecular gas fractions and mass integrated molecular gas volume densities at the high mass end (Figure 8, right panel, Figure 9, left panel), which has been noted in the discussions of most of the recent observational and theoretical papers. Independent of whether the molecular data originates from pointed observations or from deep fields, the empirical results exhibit a high mass shoulder or bump that is not present in any of the models and simulations. The same is true for the high-mass tail of the *SFR* distributions at high-z. The most recent MUFASA simulations of Davé et al. (2017) and the Illustris/TNG based work (Popping et al. 2019) do a better job (left panel of Figure 8). Given the excellent agreement between the pointed and deep field techniques, and between the different molecular mass estimators, it is unlikely that the tension lies with uncertainties in the CO-$H_2$ conversion factor, as proposed by Popping et al. (2019). The only obvious problem in the data might be a zero point issue. The S17 zero points would make things still worse (left panel of Figure 9), and the relatively good agreement between observations and theory at *z*=0 (Figures 8 & 9) would indicate a redshift dependent zero point problem. This is in principle possible, since all three techniques transfer $z=0$ calibrations to higher $z$. If the tension between predicted and observed molecular masses is the result of assumptions in the theoretical models, one possible solution might be the efficiency and mass dependence of the stellar feedback, and the recycling of wind gas related to this (Oppenheimer et al. 2010; see Sidebar).

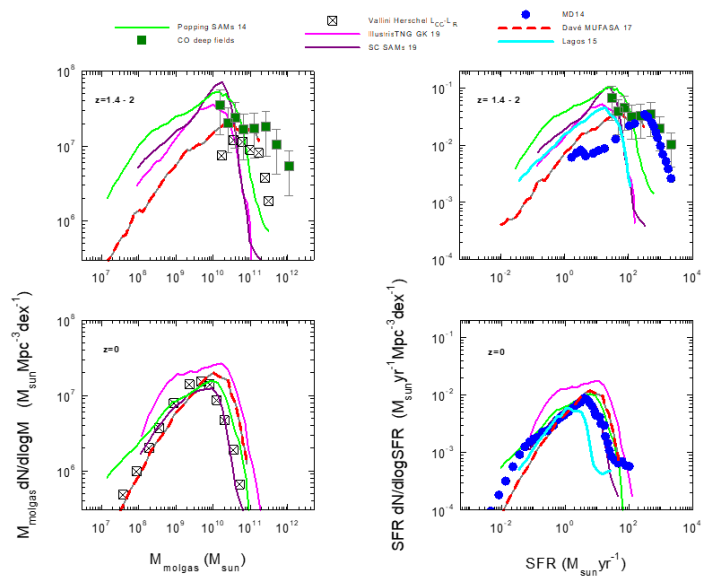

*Figure 8. Predicted and observed mass (left) and star formation rate (right) distribution functions (per co-moving volume and per logarithmic mass/SFR interval)) for redshift 0 (bottom) and 2 (top). Bottom left: The thick green and dotted red curves denote the molecular mass per logarithmic mass interval per cosmic co-moving volume ($M_{mol} \times dN/dlogM_{molgas}$) at $z=0$ from the semi-analytic models (SAMs) of Popping et al. (2014) and the MUFASA simulation of Davé et al. (2017). The continuous purple and magenta lines are the Santa Cruz SAMs and the Illustris TNG simulations results from Popping et al. (2019). The crossed black rectangles are derived from Herschel observations and turning the far-IR luminosity function into a molecular mass distribution function (Vallini et al. 2016). Top left: Same but for $z=2$. In addition the filled green squares denote the observed mass functions from current deep field observations at $z\sim2$-$2.5$ (Decarli et al. 2014, 2016, 2019, Walter et al. 2016, Riechers et al. 2019, Lenkic et al. 2019). Bottom right: $z=0$ star formation rate distributions (SFR $dN/dlogM_{molgas}$) from the same theoretical work as in the left panes, as well as from the post-processing of the EAGLE cosmological simulation (Schaye et al. 2015) by Lagos et al. (2015, cyan). The blue circles denote estimates obtained from combining the UV- and far-IR-luminosity functions of MD14, and multiplying the result by $SFR/t_{depl}=1.6 \times (1+z)^{-1}$, where we assume that a luminosity of $10^{10} L_\odot$ corresponds to a star formation rate of $1 M_\odot yr^{-1}$ (Chabrier IMF). Top right: The same but for $z\sim2$. The green squares again denote the mass density in the upper left diagram (obtained from all deep field observations) and divided by $t_{depl}=1.6 \times (1+z)^{-1}$ to turn the mass functions into star formation rate functions.*

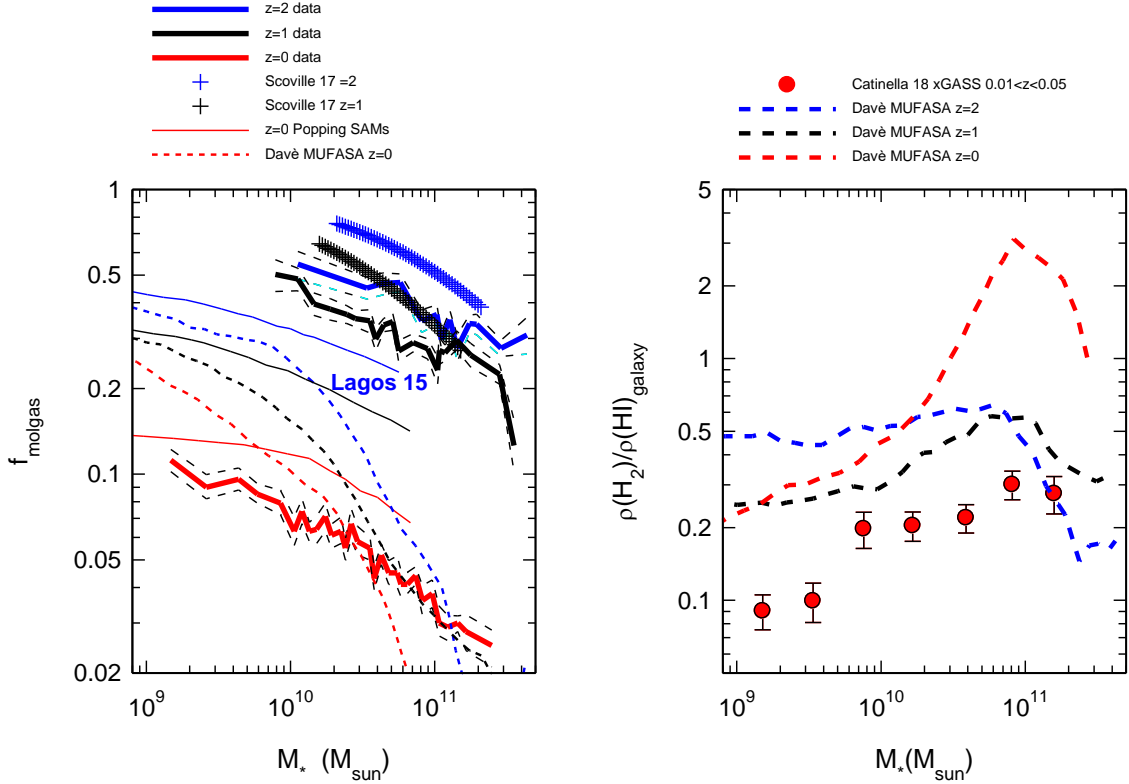

*Figure 9. Estimates of molecular gas to stellar plus molecular mass and molecular mass to atomic gas fractions, as a function of stellar mass and redshift. Left: molecular gas fractions ($f_{molgas}=(M_{molgas}/M_*)/(1+(M_{molgas}/M_*))$) as a function of stellar mass at z=0,1 and 2 in red, black and blue. The semi-analytic models of Popping et al. (2014) are continuous lines, the MUFASA simulations of Davé et al. (2017) are dotted, and the z=2 results from processing the EAGLE simulations are given as "Lagos 15". The data from the master set in this paper are thick continuous lines with black dotted lines marking the ±1σ scatter. The S17 scaling relations at z=1 and 2 are black and blue crosses. To first order, the S17 gas fractions are higher than the results from this review and in T18 because of the 50% larger CO conversion factor. Right: Intra-galaxy $H_2$ to HI ratios from the MUFASA simulations are thick dotted curves, again in the three redshift bins, as in the left panel. The observed ratios are given as a thick red line on grey circles (from Catinella et al. 2018).*

All theoretical studies predict most of the molecular gas to be associated with low mass systems (Figures 8 & 9), yet neither the pointed, nor the deep field approaches have been able to assemble statistically meaningful results for $<10^{10}$ $M_\odot$ galaxies at high-z and $<10^9$ $M_\odot$ galaxies at z=0. This failure has been discussed in detail in the literature (see references in section 2.1.2), and is plausibly closely connected to the photo-dissociation of CO ('dark gas') and low dust-to-gas ratios in low metallicity systems, making low mass systems hard to detect.

## 4.3 The Gas Regulator Model

A number of the basic results we have discussed in this review can be understood in a simple, and observationally testable, analytical framework by considering the flow of gas into galaxies, conversion into stars by in situ star formation, and the ejection of gas out of galaxies by stellar or AGN 'feedback'. This 'bathtub' (Bouché et al. 2010; Davé et al. 2012; Dekel & Mandelker 2014, Somerville & Davé 2015 ), or 'gas regulator' (Lilly et al. 2013; Peng & Maiolino 2014) model starts with the continuity equation for the gaseous and stellar matter

$$\dot{\Phi} = (1-R+\eta) \times SFR + \dot{M}_{molgas} \qquad (7),$$

where the left side of the equation is the accretion rate of gas into the galaxy, $M_{molgas}$ and $\dot{M}_{molgas}$ are the mass and the change in mass of the molecular gas in the galaxy. R is the return fraction of gas back into the galaxy by massive stars (0.4 for Chabrier IMF), and $\eta$ is the mass loading of stellar feedback as defined in section 4.2. The last term on the right side is the time derivative of the galactic molecular gas reservoir (the level of the reservoir). Simulations show that the baryonic accretion rate into the galaxy $\dot{\Phi}_{gal,baryon}$ and the dark matter accretion rate into the halo $\dot{\Phi}_{h,DM}$ can be written as

$$\dot{\Phi}_{gal,baryon} = f_{gh} \times f_{baryon} \times \dot{\Phi}_{h,DM} = 6.3 \times f_{gh} \times \left(\frac{f_{baryon}}{0.18}\right) \times \left(\frac{M_h}{10^{12} M_\odot}\right)^{1.15} \times (1+z)^{2.35} \; M_\odot yr^{-1} \quad (8),$$

(Neistein & Dekel 2008; Genel et al. 2008), for a halo mass $M_h$ at $z$, cosmic baryon to dark matter fraction $f_{baryon}$, and transport efficiency factor from halo to galaxy $f_{gh}$. Recalling that $M_{molgas}=SFR \times t_{depl}$ and $\mu=M_{molgas}/M_*$, equation (8) can then be reformulated as

$$\dot{\Phi}_{gal,baryon} = ((1-R) \times (1+\mu_{molgas}) + \eta + t_{depl} \frac{d\ln\mu_{molgas}}{dt}) \times SFR, \text{ and} \qquad (9)$$

$$\dot{M}_{molgas} = \mu_{molgas}(1-R) \times SFR + M_* \frac{d\mu_{molgas}}{dt} = \left[\tau_{depl} \frac{d\ln\mu_{molgas}}{dt} + \mu_{molgas} \times (1-R)\right] \times SFR.$$

Taking the Lilly et al. (2013) formulation of equation (1), it is then easy to calculate the gas accretion rate density and gas reservoir change required from the observed star formation rate (MD14) volume density, and the properties of the gas reservoir (Section 3), and compare these to the expected accretion rate from the simple, dark matter based estimator above. The left panel of Figure 1 shows the results for the simple (but not realistic) Ansatz $\eta$=const.

The thick black line denotes the dark matter accretion rate volume density integrated over all masses, as a function of redshift, and the thick dashed black line is the maximum gas accretion rate, for the cosmic baryon fraction and $f_{gh}$=1. Recall that this rate is an upper limit for the baryonic accretion due to the cosmic growth of the halo. For comparison, the thin dashed grey and continuous thick cyan lines are the accretion rates obtained with the above estimate from the gas regulator, for outflow mass loading $\eta$=2 and 1, respectively.

To first order the dashed black and continuous cyan lines are comparable between $z=0$ and 5. At the peak of the cosmic galaxy/star formation activity ($z\sim2$) the cosmic accretion upper limit is below the cyan line, certainly for the large mass loading factor of $\eta=2$. Again, this suggests qualitatively that recycling of the wind-ejected matter may be an important aspect of the gas evolution (Oppenheimer et al. 2010; Davé et al. 2011a, b, 2012).

Finally, the blue curve shows the change in the level of the gas regulator. In the 'ideal' regulator, the 'gas reservoir level' of the bathtub does not change, and the galaxy star formation rate is tied only to the baryonic accretion rate from the halo. At large z the gas accretion time, $t_{acc}=M_{molgas}/dM_{molgas}/dt$, is shorter than the depletion time, such that the level of the gas reservoir rises, and the galaxy is fed gas faster than it can convert to stars (Lilly et al. 2013). Starting at the $z\sim2$ peak of the star formation rate this situation reverses and the regulator becomes more ideal, although, the recycling fountains may keep the molecular gas reservoirs high at late times, thereby pushing the time of maximum molecular mass beyond that of the peak of star formation. Overall, the simple gas regulator model is remarkably effective in describing the properties and evolution of the gas and star formation rates in galaxies.

Finally, the growth of massive black holes (green line in the left panel of Figure 1) follows that of the stars closely, as pointed out many times (MD14). This suggests that co-evolution between the stellar and black components holds on average, with large fluctuations in time due to the much more time variable activity and evolution of the black hole component (Mullaney et al. 2012; Hickox et al. 2014; Delvecchio et al. 2015).

*4.3.1 Disk Instabilities, Radial Transport and Bulge Formation.* Galaxy integrated models of equilibrium growth only yield a first order picture. Spatially resolved semi-analytic models and hydro-simulations give detailed insight into the circum-galactic and sub-galactic workings of the regulator (Forbes et al. 2012, 2014; Genel et al. 2014; Hopkins et al. 2014; Lagos et al. 2015; Zolotov et al. 2015; Bower et al. 2017; Davé et al. 2017; Anglés-Alcázar et al. 2017; Krumholz et al. 2018; Pillepich et al. 2019; Nelson et al. 2019). Counter-rotating CGM streams, mergers, and global Toomre-scale instabilities in gas rich disks create gravitational torques and rapid angular momentum re-distribution at high-z. As a result, gas (and stars) are transported radially from the outer disk to a growing central bulge, on the dynamical friction, or viscous time scale, $t_{dv} \sim t_{dyn}(R_e) \times (v_c/\sigma_0)^2 \sim$ a few $10^8$ yr at $z\sim2$ (Noguchi 1999; Immeli et al. 2004; Genzel et al. 2008; Dekel, Sari & Ceverino 2009; Dekel & Krumholz 2013; Forbes et al. 2012, 2014; Zolotov et al. 2015; Rathaus & Sternberg 2016). Likewise the baryon cycle and these transport processes determine the metallicity evolution of galaxies and plausibly establish the exponential form of the stellar distribution (Davé et al. 2011b; Forbes et al 2012, 2014; Lilly et al. 2013; Rathaus & Sternberg 2016). Observational evidence for these fast transport processes and the subsequent 'compaction' (Zolotov et al. 2015) has come from studies of high-z gas kinematics (e.g. Genzel et al. 2008) and the occurrence of compact blue optical 'nuggets' (Barro et al. 2013), or compact nuclear submillimeter dust concentrations (Barro et al. 2016; Tadaki et al. 2017). These systems might be the precursors of the population of compact high-z passive galaxies (Daddi et al. 2005; Trujillo et al. 2006; van Dokkum et al. 2008; Genzel et al. 2014a; Tacchella et al. 2015). Gas rich, globally Toomre unstable, or marginally stable ($Q \leq Q_{crit} \sim O(1)$, see Section 1.1) galaxies are turbulent and geometrically thick, $(\sigma_0/v_c) \sim (h_z/R_d) \sim Q \times f_{gas}/a$, with $a \sim 1.4\text{-}1.7$ (Genzel et al. 2008; Dekel et al. 2009). For typical gas fractions at $z \sim 2$, $f_{gas} \sim 0.45$, the Toomre instability model predicts $(\sigma_0/v_c)_{z=2}$ $\sim (h_z/R_d)_{z=2} \sim 0.25\text{-}0.3$. This is in excellent agreement with the observations, including the continuous decrease of $\sigma_0$ with redshift (sections 1.4 & 3.5; Förster Schreiber et al. 2006, 2009; Kassin et al. 2012; Wisnioski et al. 2015; Übler et al. 2019). It is tempting to speculate that the highly turbulent, gas rich phase of disk galaxies can (at least in part) be identified with the thick disk in the modern Milky Way and other spiral galaxies (Gilmore et al. 1989; Bournaud et al. 2009).

In the simulations, galaxies move up and down the MS on typical time scales of about $0.4\,H(z)^{-1}$ (Tacchella et al. 2016), as a result of these perturbations and of fluctuations in the accretion rates driven by the geometry and angular momenta of the cosmic streams feeding the galaxies.

## 4.4 The Balance of $H_2$ to HI Cold Gas Components in Galaxies

We have discussed in section 3 the rapid increase of molecular gas fractions and molecular gas volume density with increasing look-back time, leveling off at $t_{lb} \sim 10\text{-}12$ Gyr ($z \sim 2\text{-}3$), and matching the same behavior of star formation rates and expected accretion rates from the halo (Figure 1). Between $z=0$ and 2 the molecular hydrogen content of the Universe more than quadruples, and gas to stellar fractions increase by an order of magnitude. Before $z \sim 2$ the growth of molecular gas and stellar mass density tracked each other. At and after the $z \sim 2$ peak gas reservoirs and fractions began to drop, massive star forming galaxies stopped growing and transitioned to the passive galaxy sequence. The right panel of Figure 1 also shows the atomic hydrogen content of galaxies (and their CGMs), as estimated from 21 HI emission observations at $z<0.4$, and from Lyman-α observations in absorption against stars and distant AGN. The cosmic atomic hydrogen density per co-moving volume, integrated over all galaxy masses, increased only 50% between $z=0$ and 2. Since at least a fraction of the hydrogen at high-z inferred from Lyα plausibly resides in the CGM, the atomic hydrogen content in the central galaxies may increase even less. Yet the cold ISM in SFGs of all masses, and especially at low redshifts and low masses, is predominantly in form of atomic hydrogen (right panel of Figure 9; Young & Scoville 1991; Saintonge et al. 2011a; Catinella et al. 2018). Catinella et al. (2018) find $log(M_2/M_{HI})=0.265 \times (logM_*-10.7)-0.62$. While the central star forming disks are dominated by molecular hydrogen, the outer disks are mostly atomic (Young & Scoville 1991; Leroy et al. 2009; Saintonge et al. 2016). These observational findings are broadly captured by the theoretical work, both in SAMs (Lagos et al. 2011; Forbes et al. 2012, 2014; Popping et al. 2014) and simulations (e.g. Lagos et al. 2015; Davé et al. 2017).

## 4.5 What Drives the $H_2$ to HI Ratio?

Hydrogen gas in the interstellar medium of galaxies is present in atomic (HI) form in diffuse clouds with low visual extinctions, and in molecular ($H_2$) form in dense optically thick regions (such as in GMCs) that are well shielded against molecular photo-dissociation. Full conversion of HI to $H_2$ is necessary for the complete incorporation of gas phase carbon into CO molecules (Jansen et al. 1995; Sternberg & Dalgarno 1995). However, this is not a sufficient condition because CO is more susceptible to photo-dissociation than is $H_2$, especially at low metallicities (Wolfire et al. 2010; Nordon & Sternberg 2016; see section 2.1.2).

In the diffuse medium, the HI can exist in two "phases", the warm ~$8 \times 10^3$ K neutral medium (WNM) in which cooling is dominated by a combination of Lyα emission and, depending on metallicity, also electron recombination onto small grains and PAHs, and the cool ~100 K neutral medium (CNM), where the energy losses are via metal fine-structure line emissions, especially the $C^+$ 158 μm line (Field et al. 1969; Draine 1978; Wolfire et al. 2003; Bialy & Sternberg 2019). Heating is by far-UV absorption and then ejection of electrons from dust grains (photoelectric effect), cosmic ray ionization of the hydrogen atoms, and the dissipation of turbulence. In general, the HI is pure WNM at sufficiently low volume densities where cooling is inefficient, and pure CNM at high densities where the collision rates are rapid. Depending on the heating rates and the gas phase metallicities and dust abundances, the two phases can coexist for a narrow range of thermal pressures. CNM condensations are then embedded within an enveloping WNM, observable as narrow 21 cm CNM emission or absorption features across broad WNM emission profiles (e.g., Dickey & Brinks 1993; Heiles & Troland 2003; Stanimirović et al. 2014; Warren et al. 2012). The HI gas in galaxies may often be self-regulated to a multi-phased WNM/CNM state by a feedback loop (Krumholz et al. 2009; Ostriker et al. 2010; see also Schaye 2004). For a given galaxy disk pressure, as set for example by the overlying weight of the ISM, and assuming that star-formation requires the presence of CNM, the star-formation rate will adjust such that the HI becomes multi-phased. If the *SFR* is too large, the elevated heating rates will drive the gas to WNM thereby reducing the *SFR* and heating rates, and enabling conversion back to CNM.

In dense optically thick clouds, the atomic to molecular (HI-to-$H_2$) transition is controlled by the balance between $H_2$ formation on the surfaces of dust grains versus $H_2$ destruction by far-UV photo-dissociation. With increasing cloud depth the $H_2$ photo-dissociation rates are reduced by a combination of dust opacity, and $H_2$ absorption line opacity in the Lyman and Werner (LW) molecular band systems through which photo-dissociation occurs. The $H_2$ formation efficiency, as well as the dust absorption opacity, both depend on the dust to gas ratio or the metallicity of the gas. The $H_2$ formation rate is proportional to the gas density, and for a sufficiently weak radiation field, the HI-to-$H_2$ conversion is controlled by $H_2$ absorption line "self-shielding". In this weak field regime the conversion point occurs at a very low dust optical depth, and most of the photo-dissociated HI is present (in trace amounts) inside the molecular zones. For sufficiently intense radiation fields, the photo-dissociation rate is attenuated by the dust opacity associated with large HI column densities. In this strong field regime most of the HI is present as a fully photo-dissociated surface layer surrounding an internal molecular zone. Thus, the primary parameters controlling the HI-to-$H_2$ transition are the gas density (or pressure), the intensity of the free-space far-UV field, and the metallicity which determines the $H_2$ formation rate coefficient and dust absorption opacity. In the fully shielded molecular zones heating is inefficient and the gas becomes cold (<50 K).

Cloud structures, in terms of both density inhomogeneities and velocity distributions (e.g. as imprinted by turbulent motions) are a major complication because of the effects on the shielding and radiative transfer properties of the gas. Treatments of the HI-to-$H_2$ conversion in hydro-dynamical simulations (cosmological or zoom-in) and in semi-analytic models therefore require the adoption of (possibly quite crude) sub-grid recipes, usually based on simple analytic "single cloud" formulae (Sternberg 1988, Elmegreen 1993; Blitz & Rosolowsky 2006; Krumholz et al. 2008, Gnedin & Kravtsov 2011; McKee & Krumholz 2010; Sternberg et al. 2014). Such recipes have been used in computations in which star-formation is assumed to require the presence of $H_2$ (Fu et al. 2010; Christensen et al. 2012; Kuhlen et al. 2012; Thompson et al. 2014; Popping et al. 2014, 2015; Davé et al. 2016; Xie et al. 2017), or simply for partitioning the cold gas between HI and $H_2$ (Obreschkow et al. 2009; Lagos et al. 2011; Bekki 2013; Lagos et al. 2015; Marinacci et al. 2017; Diemer et al. 2018). Simulations have also been carried out in which the (non-equilibrium) $H_2$ formation/destruction chemistry is computed "on the fly" using subgrid prescriptions for the photo-dissociating radiation fields and cold gas densities (Robertson & Kravtsov 2008; Hu et al. 2016; Luppi et al. 2018; Tomasseti et al. 2015; Nickerson et al. 2018, 2019).

A simple expression for the $H_2$ gas fraction, $f_{H2}$, in optically thick uniform density one-dimensional (plane parallel or spherical) clouds with a total hydrogen (atomic plus molecular) column density, $N$ (cm$^{-2}$) is (Sternberg et al. 2014)

$$f_{H2} = 1 - \frac{N_{HI}}{N} = 1 - \frac{1.6 \sigma_g^{-1} \times \ln(\alpha_{df} G / 3.2 + 1)}{N} \qquad (10),$$

where $\sigma_g$ is the metallicity dependent dust absorption cross section per hydrogen nucleus, and $N_{HI}$ is the total HI column density is produced by photo-dissociation is for two sided illumination of a plane-parallel slab, or equivalently isotropic irradiation of a sphere (and neglecting cosmic-ray destruction of the $H_2$). Equation (10) is for optically thick clouds for which $N > N_{HI}$ by definition. In this expression, $\alpha_{df} = D/R'n$ is the ratio of the free-space $H_2$ dissociation rate, $D$, to the molecular formation rate, $R'n$, where $R'$ is the grain-surface formation rate coefficient of $H_2$, and $n$ is the total (atomic plus molecular) gas density. $G$ is a metallicity dependent cloud-averaged $H_2$ self-shielding factor. The product, $\alpha_{df}G$, is the primary dimensionless parameter in the problem, and is given by

$$\alpha_{df} G = 1.54 \times \left(\frac{\sigma_g}{1.9 \times 10^{-21} cm^{-2}}\right) \times \left(\frac{F_{UV}}{2 \times 10^7 cm^{-2} s^{-1}}\right) \times$$
$$\times \left(\frac{R'}{3 \times 10^{-17} cm^3 s^{-1}}\right)^{-1} \times \left(\frac{n}{10^2 \, cm^{-3}}\right)^{-1} \times \left(1 + [2.64 Z']^{0.5}\right)^{-1} \qquad (11).$$

In this expression, $F_{UV}$ is the flux of far-UV photons in the Lyman-Werner bands, and $Z'$ is the metallicity in units of solar metallicity. The above equation is quite general and provides insight into the behavior of the $H_2$ gas fraction in terms of the above parameters. For $\alpha_{df}G < 1$, $H_2$ self-shielding controls the atomic to molecular transition. For $\alpha_{df}G > 1$, the transition is governed by dust absorption of the radiation. For any value of $\alpha_{df}G$, $N_{HI}$ is inversely proportional to $\sigma_g$. This means that the atomic column at the surface of the cloud is larger for smaller metallicities, and the total cloud column $N=2N_{HI}$ required for a given H$_2$ mass fraction of is also correspondingly larger. If the self-regulation Ansatz for multi-phased phased HI gas is assumed (Krumholz et al. 2009) then $\alpha_{df}G \sim 2$ for the 'cold neutral medium' component (CNM), or $N_{HI}=0.7/\sigma_g$. With this assumption, $f_{H2}=0.5$ occurs for clouds with $N=1.4/\sigma_g$, corresponding to a mass density of $N_{crit} = 12/Z'$ M$_\odot$ pc$^{-2}$, for $\sigma_g$ scaling linearly with metallicity $Z'$, in units of the solar metallicity.

A second estimate of this critical column density comes from observations of the HI and H$_2$ gas content of nearby galaxies. Blitz & Rosolowsky (2006) find that the *H$_2$/HI* ratio varies as $P_e^{0.9}$ where $P_e$ is the pressure. Since $P_e \propto \Sigma_{gas}^2$, they find $H_2/HI \sim (\Sigma_H/45\ M_\odot pc^{-2})^{1.8}$. The critical column density at which *H$_2$/HI=1* (i.e. $f_{H2}=0.5$) is in rough agreement with the photo-dissociation argument. This critical column density is also in good agreement with a kink in the KS-plane of $\Sigma_{gas}$–$\Sigma_{SFR}$ below which the slope of the relation becomes much steeper than $N_{KS}$=1-1.4 (Wong & Blitz 2002; Kennicutt 2008; Bigiel et al. 2008; Kennicutt & Evans 2012).

The observed evolution of the cosmic molecular hydrogen to atomic hydrogen ratio is matched quite well by the theoretical work, especially the most recent simulations (right panel of Figure 1; Davé et al. 2017), showing that the above modeling techniques probably do a satisfactory job for the global evolution of the gas reservoirs.

## 5. Starburst Galaxies

So far, our review has focused on MS galaxies that grow in quasi-equilibrium at the average gas accretion rate (equation (8)). In this case the growth time of the galaxy is directly proportional to the growth time of the dark matter halo, $t_{DM} = M_{DM}/\dot{M}_{DM} \approx 2.1 \times ([1+z]/3)^{-2.3}$ Gyr for a halo mass of $10^{12}$ M$_\odot$. The mass dependence of $t_{DM}$ is very shallow, $t_{DM} \propto M_{DM}^{-0.15}$. The growth time of the stellar component is $t_* = M_*/(SFR \times (1-R)) = 0.53 \times ((1+z)/3)^\gamma$ Gyr, where $\gamma$=-3 for z=0-2, and $\gamma$=-1.67 for z>2 (Lilly et al. 2013). Note that the stellar component grows faster by a factor of ~2 (z~0) to 4 (z~2) than the DM-halo, due to the effects of feedback, reducing the numerator at early times Since the mass loading for momentum driven stellar feedback is expected to scale as $\eta \sim v_c^{-1..2}$ (Davé et al. 2012), smaller galaxies have a shallower potential and more powerful winds.

A $10^{12}$ M$_\odot$ halo will experience a binary major merger (<3:1 mass ratio) every $t_{major\ merger}$ ~ 4.4 Gyrs $\times ((1+z)/3)^{-2.1}$ (Neistein & Dekel 2008; Genel et al. 2009, 2010; Fakhouri & Ma 2009). This is comparable to the time scale when two nodes of the cosmic web merge, and therefore, the angular momentum of the gas coming into the galaxy changes drastically (Dekel, priv. comm.). Once that happens, the two galaxies will experience strong gravitational torques that drive gas inward, compress it in the central regions and trigger an elevated starburst activity (Mihos & Hernquist 1996). Minor mergers with mass ratios 3:1 to 10:1 are several times more frequent but naturally have a lower efficacy in triggering bursts of elevated star formation. Interaction driven starbursts last for typically a few dynamical times of the merger, ~$10^8$ years << $t_{depl}$(MS). The phenomenon of interaction driven starburst activity was initially discovered in nearby dusty galaxies, such as M82 and NGC 253 (e.g. Rieke et al. 1980, 1993; Heckman et al. 1998). Sanders et al. (1988), Rowan-Robinson & Crawford (1989) and others found extreme versions of dusty, infrared bright starbursts in the IRAS survey, ULIRGs (e.g. Sanders & Mirabel 1996). ULIRGs form stars at 100-300 $M_\odot yr^{-1}$, which is about two orders of magnitude above the $z$=0 MS at their stellar mass of a few $10^{10}$ $M_\odot$. ULIRGs are invariably mid- to late stage, major mergers of moderately massive and gas-rich disks, with star formation and AGN contributing to the total luminosities (Genzel et al. 1998).

A possibly analogous, highly obscured star-forming population at high-z (submillimeter galaxies, or SMGs) was discovered through sub-millimeter observations with the SCUBA sub-millimeter bolometer (e.g. Smail et al. 1997; Blain et al. 2003). SMGs are often described as scaled up, high-z versions of ULIRGs, with SFRs of 300-2000 $M_\odot$ yr$^{-1}$. Many SMGs are indeed compact, merger driven starbursts much above the $z$~1-3 MS (e.g. Tacconi et al. 2006, 2008; Engel et al. 2010; Bothwell et al. 2013; Casey et al. 2014; Wiklind et al. 2014; Hodge et al. 2016). Others are very massive MS galaxies at the tip of the mass-SFR relation at $z$~2 (Michalowski et al. 2012). The widest Herschel survey, ATLAS, uncovered a more extreme, rare population of hyper-luminous galaxies with $L_{IR}$ > $10^{13}$ $L_\odot$ ( Eales et al. 2010). Many of these are extremely luminous as a result of amplification through strong gravitational lensing (e.g. Frayer et al. 2011; Valtchanov et al. 2011).

ULIRGs and SMGs are >100 times rarer than the overall *log ($M_*/M_\odot$)>10* MS population at the same redshift. Rodighiero et al. (2011, 2015) have shown conclusively that these starbursts make up 6-15 % of the cosmic SFR density at $z$=0-2. The star formation and ISM physics in these objects is extreme and very different from the typical MS systems (e.g. Downes & Eckart 2007; Scoville et al. 2017; Hodge et al. 2016), and we have included extreme objects in assembling scaling relations as a function of main sequence offset (Section 3 and Table 1). The scope of this review does not allow us to discuss these interesting outlier objects in more detail, however, and there are recent comprehensive reviews on this topic in the literature (e.g. Casey et al. 2014; Wiklind et al. 2014).

# 6. Summary & Concluding Remarks

**Baryonic Accretion as the main Driver of Cosmic Star/Galaxy Formation.** We have emphasized the wealth and increasing robustness of new observations of the cosmic evolution of molecular (and atomic gas) in galaxies obtained during the last decade. The new observations motivate a model framework for the baryon cycle, star-formation, and the cosmic growth of galaxies. In the basic picture, most galaxy evolution at and since the maximum of cosmic star formation 10 Gyrs ago took place in rotationally supported disks, which mostly grew by semi-continuous baryonic gas accretion (and minor mergers) from the cosmic web and internal, local star formation in dense, dusty molecular gas. Throughout this epoch, galaxy formation efficiency is ~10-20% (of the cosmic baryons) within a halo mass range from $10^{11}$ to $10^{12.5}$ $M_\odot$, but drastically reduced outside this range (Bouché et al. 2010; Davé et al. 2012; Behroozi et al. 2013). The low formation efficiency of galaxies at the low mass end is likely caused by stellar feedback and photo-ionization at high-z. At the high mass end, at and above the Schechter mass (halo masses >$10^{12} M_\odot$) the low efficiency is probably due to a combination of inefficient gas cooling in the halo and AGN feedback (Rees & Ostriker 1978; Dekel & Birnboim 2006; Croton et al. 2006; Bouché et al. 2010; Nelson et al. 2019). The prominent maximum in cosmic star and galaxy formation rate ~10 Gyrs ago ($z$~1-3; MD14; Lilly et al. 1996; Madau et al. 1996; Steidel et al. 1996) was thus due primarily to higher gas accretion rates, and secondarily to more efficient star formation in globally unstable, gas rich systems.

**The Drivers of the Cosmic Baryon Cycle**. Many of the galaxy integrated stellar, star formation, and gas properties, and their mutual scaling relations, are captured in various versions of a simple equilibrium gas regulator model, which exploits the mass continuity rate equation, either in a galaxy integrated, or spatially resolved fashion (Bouché et al. 2010; Davé et al. 2012; Lilly et al. 2013; Dekel & Mandelker 2014; Tacchella et al. 2016; Forbes et al. 2012, 2014; Rathaus & Sternberg 2016).

Figure 10 captures and summarizes the key aspects of this model by comparing the time scales (or rates) of this baryonic gas regulator system. As discussed in section 3, the depletion time on the MS, controlling the maximum rate of forming stars in galaxies, has a time/redshift dependence plausibly tied to the dynamical time, or Toomre time, in a $Q<1$ globally unstable disk, which in turn is tied to the Hubble time at $z$, $H(z)^{-1}$. The empirically determined time for doubling the stellar component, $t_*$, is ~3-4 times smaller (or the growth rate 3-4 time faster) than the corresponding growth time of the dark matter halo in which the galaxy is embedded. At small masses the efficiency of galaxy formation is low, due to very efficient stellar feedback.

For $z < z_{peak}$ ~2, the gas consumption time is smaller than $t_*$, and the growth of the galaxy is tied to the cosmic baryon accretion rate, plus consumption of the gas reservoir, which is initially substantial but then used up at $z<0.2$. This means that in the local Universe, where $t_*$ is larger than the Hubble time, all but the smallest galaxies stop growing (are 'starved'), with the exception of gas returned by stars, and if HI from the outer disks can be brought into the inner regions. However, for $z >z_{peak}$, the gas consumption time was larger than $t_*$. In this limit, the SFR is determined by the depletion time, and a significant fraction of the baryonic accretion builds up the internal gas reservoir (Lilly et al. 2013).

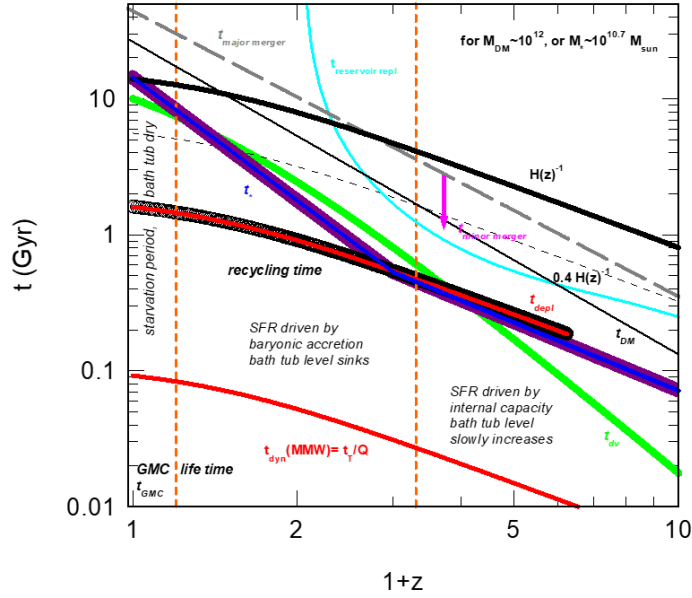

*Figure 10.* Summary of the various time scales affecting the galactic baryon cycle, plotted as a function of redshift (see sections 4, 5 & 6), using the relations summarized in this paper, and applicable for a $10^{12}$ $M_\odot$ halo. These time scales are (approximately from shortest to longest) the life time of a GMC $t_{GMC}$, the dynamical time ($t_{dyn}$) or Toomre time ($t_T$) of a typical massive galaxy with angular momentum parameter $\lambda \sim 0.037$ (red line), the dynamical friction or viscous time scale ($t_{dv}$) for radial gas transport inside the galaxy (green), the stellar growth time ($t_*$, blue), the molecular gas depletion time scale ($t_{depl}$, thick red-black), the growth time of the dark matter halo ($t_{DM}$, thin black), the molecular gas reservoir replenishment time ($t_{reservoir-rep}$, cyan), the typical time for up-down movement in the MS-plane $0.4\ H(z)^{-1}$ (thin grey dashed), the time between major mergers ($t_{major-merger}$, thick grey dashed), the time for minor mergers ($t_{minor-merger}$, magenta arrow) is down-ward from that line, and the Hubble time $H(z)^{-1}$ (thick black line). The epoch of re-cycling is at $z \sim 1$-$2$ with a typical time of 0.5 Gyr. The red vertical dashed lines indicate the times in the equilibrium growth model when the SFR is driven by the available reservoir (at high redshifts), when it is driven by accretion (near the peak of cosmic SF), and when the reservoir is running dry (at low redshifts).

Galaxies are richest in molecular gas at redshifts somewhat smaller than the epoch of maximum cosmic star formation rate. A possible explanation, is that mass loading of galactic winds at early times and in small galaxies is high. Much of this 'recycled' gas returns at later cosmic time ($z \sim 1$-$2$) into more massive galaxies, and increases their gas reservoirs and their star formation rates above and beyond what would be expected from baryonic halo accretion (Oppenheimer et al. 2010). At all redshifts, diffuse gas accretion and minor mergers dominate the growth rate from major mergers by a substantial factor (e.g. Genel et al. 2008, 2009, 2010; Fakhouri & Ma 2008, 2009; Dekel et al. 2009). Minor mergers can play a significant role at $z > 2.5$-$5$, but not at and below the peak of cosmic galaxy formation (Lilly et al. 2013 (their Figure 4); and Bouché et al. 2010; Davé et al. 2012; Dekel & Mandelker 2014).

Galaxies move up and down in the stellar mass–star formation rate plane on a time scale of a fraction of the Hubble time (Tacchella et al. 2016), due to the combination of the perturbation by the disk instabilities, mergers and changes in the orientation, geometry and angular momentum of the cosmic streams 'feeding' the galaxy. At $z>1.5$, global disk instabilities and minor mergers transport gas (and stars) radially inward on the viscous time scale, which is only ~1 Gyr or less, because of the high gas fractions and gravitationally driven turbulence. This scale is comparable to the depletion and baryonic times, resulting in efficient buildup of central star-forming bulges.

**Measurement Tools and Methods**. We have addressed how one can most reliably determine the molecular gas content of galaxies, and how one can most efficiently and with minimal bias survey the molecular gas content of a cosmic volume. Far-infrared and submillimeter dust emission are now available as tracers to check, and in some instances replace the classical CO luminosity method by being more time efficient (Bolatto et al 2013; Scoville et al. 2014; S17). We have shown that apart from zero-point offsets the three techniques give similar scaling relations, mitigating concerns about 'uncertain conversion factors'. 'Pointed' molecular gas observations of individual galaxies (or stacks) galaxies selected from the multi-band deep imaging surveys and 'deep field scanning' observations yield comparable evolutions of the molecular gas volume density as a function of redshift, or redshift and mass.

One major problem of the CO method, and probably also of the dust methods, is that emissivity (per mass) depends inversely on metallicity, such that it is hard or impossible to detect low mass galaxies with substantially sub-solar metallicity, and still harder to determine their gas content quantitatively. The result is that there are so far few CO or dust detections of low-mass galaxies ($log(M_*/M_\odot)<<10$) at high-z. It is not obvious how substantial progress can be made on this aspect. Another unknown is whether an important ISM component is completely missing or overlooked with the current techniques, such as gas/dust at very low temperature.

**Comparison of Observations and Theoretical Work**. We have compared the empirical data to a range of semi-analytic and hydro-simulation based, theoretical papers. Broadly, the theoretical (typically calibrated on $z$=0 results) and observational work of galaxy and mass integrated molecular and atomic hydrogen content per cosmic volume agree reasonably well. Both find that compared to the atomic gas, molecular gas is sub-dominant at all redshifts, but it evolves rapidly with redshift, peaking near but at somewhat lower redshift than the $z$~1-3 cosmic star formation rate maximum. Inspecting in more detail the mass functions, or gas to stellar mass ratios, as a function of redshift, the theoretical work predicts z~1-2 massive galaxies near the Schechter mass, to be less molecular-gas rich than found from observations by a factor of several, while there appears to be no disagreement at $z$=0. All SAMs and simulation-based work predict the majority of the molecular gas to be associated with lower mass galaxies. Given the difficulty to detect molecular gas in these lower mass systems at $z$~1-3 with any of the three observational techniques, it is not clear whether there is also a tension at the low mass tail. In addition, the same result is obtained when the cosmic star formation rate luminosity functions (MD14) are converted to molecular mass functions with an adopted depletion time scaling relation (Figure 8). This overall consistency is encouraging. The complexity of the combined effects of recycling and sub-galactic processes may be at the core of the tensions between simulations and observations.

**Outlook to the Next Decade**. In the opinion of the authors, there are likely three main avenues of research in which very fruitful progress can be expected in the next decade.

- One is the **extension to higher redshift**. Will the physical phenomena we encounter on this journey primarily be characterized by straightforward extrapolation of the gas regulator system to earlier times, where rates are faster but the gas-to-star-converters are on average smaller and less efficient, in terms of lower in-take and consumption, as well as in higher mass loaded outflows (Figure 10)? The hydro-simulations suggest that prior to the cosmic star formation rate peak, at $z$~3-6, galaxies were much less settled in rotationally stabilized disks and much more perturbed by mergers, resulting in a more irregular growth, akin to that of lower mass and dwarf galaxies at later cosmic time (e.g. Wetzel et al. 2016; Feldmann et al. 2019, Simons et al. 2017). The James Webb Space Telescope (JWST) has a currently scheduled launch date in 2021, shortly after the publication of this paper, and will provide a plethora of high-$z$ SFG targets for subsequent studies of the gas and star formation evolution at the earliest epochs of galaxy formation. The combination of ALMA and JWST in particular will result in much progress, but a **significant** and **coordinated** investment of time of both facilities will be necessary for robust statistical results.
- The second obvious avenue will be **spatially resolved studies**, in particular to explore the gas regulation and consumption processes on cloud and star cluster scales at $z$~1-3. A clear lesson of the last decade has been that sub-galactic scale processes can 'radiate back out' and determine star formation rates, gas densities, outflows, metallicities and perhaps even dark matter distributions on the scales of the outer disks and the CGM. Again, JWST, ALMA/NOEMA and the 30m-class optical/IR telescopes will be ideally suited for this task. Important work will focus on the radial transport and build-up of early gas-rich bulges that are predicted in the model of globally unstable disks, and for which first observations are becoming available.
- The third avenue will be the **connection of the molecular gas processes and kinematics to the incoming and outgoing ionized gas in the CGM**, and the atomic gas in the outer disk. It is highly desirable to image the baryonic gas of selected galaxies in all phases (molecular, atomic, ionized and very hot), initially in modest-redshift galaxies, where such imagery is possible with the pre-SKA precursors, and where CGM tracers can be imaged in emission. **Careful coordination between the relevant large telescopes will be required for such ambitious multi-wavelength work to succeed and yield statistically robust answers**.

Naturally, the **universal baryon-cycle-model** may not always be applicable. We have mentioned the $z$>3 epoch where this model may fail and be replaced by a much more random set of consecutive merger events. There are also cases of star formation in 'extreme environments', such as in tidal tail dwarfs, or in jet-induced star formation, or some recent observations of molecular gas in the CGM, where the model may also be inapplicable. One might ask whether gas-rich but star formation poor galaxies exist. If so, deep multi-band surveys in the submillimeter continuum (as extinction free indicator of SFR and dust mass) may be required to identify candidates.

*Acknowledgements:* We are grateful to a number of colleagues who have given us comments, advice and input on various versions of this draft. In particular, we thank Alberto Bolatto, Andi Burkert, Avishai Dekel, Rob Kennicutt, Adam Leroy, Chris McKee Thorsten Naab, Alvio Renzini, and Andreas Schruba for extremely valuable inputs and/or comments. We thank the members of the PHIBSS, xCOLDGASS, SINS and KMOS[3D] teams who have been deeply involved in throughout the years in various aspects of the research discussed in this review, especially Alberto Bolatto, Francoise Combes, Alessandra Contursi, Natascha Förster Schreiber, Santiago Garcia-Burillo, Rodrigo Herrera-Camus, Minju Lee, Dieter Lutz, Roberto Neri, Amelie Saintonge and Hannah Übler. Finally, we thank the entire staff of IRAM for their many years of support of the authors' science, which made much of the work in this review possible.

# 7. Literature Cited

Abdo AA, Ackermann M, Ajello M, et al. 2010. Ap. J. 710:133
Ali ZS, Parsons AR, Zheng H, et al. 2015. Ap. J. 809:61
Alves J, Lombardi M & Lada CJ 2007. Astron. Astrophys. 462:17
André P, Men'shchikov A, Bontemps S, et al. 2010. Astron. Astrophys. 518:102
Anglés-Alcázar D, Faucher-Giguère C-A, Kereš D, et al. 2017. MNRAS 470:4698
Aravena M, Decarli R, Gónzalez-López J, et al. 2019. arXiv:1903.09162
Aravena M, Decarli R, Walter F, et al. 2016. Ap. J. 833:68
Aravena M, Hodge JA, Wagg J, et al. 2014. MNRAS 442:558
Arimoto N, Sofue Y & Tsujimoto T 1996. Publ. Ast. Soc. Jap. 48:275
Armus L, Mazzarella JM, Evans AS, et al. 2009. Publ. Ast. Soc. Pac. 121:559
Asplund M, Grevesse N, Sauval AJ, Allende Prieto C & Kiselman D 2004. Astron. Astrophys. 417:751
Barro G, Faber SM, Pérez-González PG, et al. 2013. Ap. J. 765:104
Barro G, Kriek M, Pérez-González PG, et al. 2016. Ap. J. 827:32
Bauermeister A, Blitz L, Bolatto A, et al. 2013. Ap. J. 763:64
Beckwith SVW, Stiavelli M, Koekemoer AM, et al. 2006. Astron. J. 132:1729
Behrendt M, Burkert A & Schartmann M 2015. MNRAS 448:1007
Behroozi PS, Conroy C, & Wechsler RH 2010. Ap. J. 717:379
Behroozi PS, Wechsler RH & Conroy C 2013. Ap. J. 770:57
Bekki K 2013. MNRAS 432:2298
Bell EF, van der Wel A, Papovich C, et al. 2012. Ap. J. 753:167
Berta S, Lutz D, Genzel R, Förster Schreiber NM & Tacconi LJ 2016. Astron. Astrophys. 587:73
Bertemes C, Wuyts S, Lutz D., et al. 2018. MNRAS 478:1442
Béthermin M, Daddi E, Magdis G, et al. 2015. Astron. Astrophys. 573:113
Bezanson R, van der Wel A, Straatman C, et al. 2018. Ap. J. 868:36
Bialy S & Sternberg A 2015. MNRAS 450:4424
Bialy S & Sternberg A 2019. Ap. J. 881:160
Bigiel F, Leroy A, Walter F, et al. 2008. Astron. J. 136:2846

Bigiel F, Leroy AK, Walter F, et al. 2011. Ap. J. 730:13
Binney J & Tremaine S 2008. Galactic Dynamics: Second Edition Princeton: Princeton University Press
Bisbas TG, Papadopoulos PP & Viti S 2015. Ap. J. 803:37
Bisbas TG, van Dishoeck EF, Papadopoulos PP, et al. 2017. Ap. J. 839:90
Blain AW, Barnard VE & Chapman S 2003. MNRAS 338:733
Bolatto AD, Leroy AK, Rosolowsky E, et al. 2008. Ap. J. 686:948
Bolatto AD, Wolfire M & Leroy AK 2013. Annu. Rev. Astron. Astrophys. 51:207
Bolatto AD, Wong T, Utomo D, et al. 2017. Ap. J. 846:159
Boselli A, Eales S, Cortese L, et al. 2010. Pub. Ast. Soc. Pac. 122:261
Boselli A & Gavazzi G 2014. Astron. Astrophys. Rev. 22:74
Bothwell MS, Smail I, Chapman SC, et al. 2013. MNRAS 429:3047
Bouché N, Dekel A, Genzel R, et al. 2010. Ap. J. 718:1001
Bournaud F & Elmegreen BG 2009. Ap. J. 694:158
Bournaud F, Elmegreen BG & Martig M 2009. Ap. J. Lett. 707:L1
Bouwens RJ, Bradley L, Zitrin A, et al. 2014. Ap. J. 795:126
Bower RG, Schaye J, Frenk CS, et al. 2017. MNRAS 465:32
Brammer GB, van Dokkum PG, Franx M, et al. 2012. Ap. J. Suppl. 200:13
Braun R 2012. Ap. J. 749:87
Breysse PC & Rahman M 2017. MNRAS 468:741
Brinchmann J, Charlot S, White SDM, et al. 2004. MNRAS 351:1151
Brusa M, Bongiorno A, Cresci G, et al. 2015. MNRAS 446:2394
Burkert A, Förster Schreiber NM, Genzel R, et al. 2016. Ap. J. 826:214
Carilli CL & Walter F 2013. Annu. Rev. Astron. Astrophys. 51:105
Casey CM, Narayanan D & Cooray A. 2014. Phys. Rep. 541:45
Catinella B, Saintonge A, Janowiecki S, et al. 2018. MNRAS 476:875
Chabrier G 2003. Pub. Ast. Soc. Pac. 115:763
Christensen C, Quinn T, Governato F, Stilp, A, Shen S & Wadsley, J 2012. MNRAS 425:3058
Combes F 2018. Astron. Astrophys. Rev. 26:5
Combes F, García-Burillo S, Braine J, et al. 2011. Astron. Astrophys. 528:124
Combes F, Garcıa-Burillo S, Braine J, et al. 2013. Astron. Astrophys. 550:41
Cowie LL, González-López J, Barger AJ, et al. 2018. Ap. J. 865:106
Croton DJ, Springel V, White SDM, et al. 2006. MNRAS 365:11
Crutcher, R.M. 2012. Annu. Rev. Astron. Astrophys. 50:29
Daddi E, Bournaud F, Walter F, et al. 2010a. Ap. J. 713:686
Daddi E, Dannerbauer H, Liu D, et al. 2015. Astron. Astrophys. 577:46
Daddi E, Dickinson M, Morrison G, et al. 2007. Ap. J. 670:156
Daddi E, Elbaz D, Walter F, et al. 2010b. Ap. J. Lett. 714:L118
Daddi E, Renzini A, Pirzkal N, et al. 2005 Ap. J. 626:680
Dale DA & Helou G 2002. Ap. J. 576:159
Dame TM, Hartmann D & Thaddeus P 2001. Ap. J. 547:792
Dannerbauer H, Daddi E, Riechers DA, et al. 2009. Ap. J. Lett. 698:L178
Darvish B, Scoville NZ, Martin C, et al. 2018. Ap. J. 860:111
Davé R, Anglés-Alcázar D, Narayanan D, et al. 2019. MNRAS 486:2827
Davé R, Finlator K & Oppenheimer BD 2011b. MNRAS 416:1354

Davé R, Finlator K & Oppenheimer BD 2012. MNRAS 421:98
Davé R, Oppenheimer BD & Finlator K 2011a. MNRAS 415:11
Davé R, Thompson R & Hopkins PF 2016. MNRAS 462:3265
Davé R, Rafieferantsoa MH, Thompson RJ & Hopkins PF 2017. MNRAS 467:115
DeBoer DR, Parsons AR, Aguirre JE, et al. 2017. Publ. Astron. Soc. Pac. 129:974
Decarli R, Walter F, Aravena M, et al. 2016. Ap. J. 833:70
Decarli R, Walter F, Carilli C, et al. 2014. Ap. J. 782:78
Decarli R, Walter F, Gónzalez-López J, et al. 2019. ArXiv: 1903.09164
Dickman RL, Snell RL & Schloerb FP 1986. Ap. J. 309:326
Dekel A & Birnboim Y 2006. MNRAS 368:2
Dekel A, Birnboim Y, Engel G, et al. 2009. Nature 457:451
Dekel A & Krumholz MR 2013. MNRAS 432:455
Dekel A & Mandelker N 2014. MNRAS 444:2071
Dekel A, Sari R & Ceverino D 2009. Ap. J. 703:785
Delvecchio I, Lutz D, Berta S, et al. 2015. MNRAS 449:373
Dickey JM & Brinks E 1993. Ap. J. 405:153
Diemer B, Stevens ARH, Forbes JC, et al. 2018. Ap. J. Suppl. 238:33
Dobbs CL, Burkert A & Pringle JE 2011. MNRAS 413:2935
Dobbs CL & Pringle JE 2013. MNRAS 432:653
Downes D & Eckart A 2007. Astron. Astrophys. 468:57
Downes D & Solomon PM 1998. Ap. J. 507:615
Draine BT 1978 Ap. J. Suppl. 36:595
Draine BT, Dale DA, Bendo G, et al. 2007. Ap. J. 663:866
Draine BT & Li A 2007. Ap. J. 657:810
Draine, B.T. 1978. Ap. J. Suppl. 36:595
Dunlop JS, McLure RJ, Biggs AD, et al. 2017. MNRAS 466:861
Eales S, de Vis P, Smith MWL, et al. 2017. MNRAS 465:3125
Eales S, Dunne L, Clements D, et al. 2010. Publ. Astron. Soc. Pac. 122:499
Eales S, Fullard A, Allen M, et al. 2015. MNRAS 452:3489
Elbaz D, Daddi E, Le Borgne D, et al. 2007. Astron. Astrophys. 468:33
Elbaz D, Dickinson M, Hwang HS, et al. 2011. Astron. Astrophys. 533:119
Elmegreen BG 1993. Ap. J. 411:170
Elmegreen BG 1997. Rev. Mex. Astr. Astrof. 6:165
Elmegreen BG 2007. Ap. J. 668:1064
Elmegreen BG & Scalo J 2004. Annu. Rev. Astron. Astrophys. 42:211
Engel H, Tacconi LJ, Davies RI, et al. 2010. Ap. J. 724:233
Epinat B, Tasca L, Amram P, et al. 2012. Astron. Astrophys. 539:92
Escala A 2011. Ap. J. 735:56
Faber SM, Willmer C, Wolf C, et al. 2007. Ap. J. 665:265
Faerman Y, Sternberg A & McKee CF 2017 Ap. J. 835:52
Fakhouri O & Ma C-P 2008. MNRAS 386:577
Fakhouri O & Ma C-P 2009. MNRAS 394:1825
Fall SM & Romanowsky AJ 2013. Ap. J. 769:26
Faucher-Giguère C-A & Quataert E 2012. MNRAS 425:605
Feldmann R, Faucher-Giguère C-A & Kereš D 2019. Ap. J. 871:21
Feldmann R, Gnedin NY & Kravtsov AV 2012a. Ap. J. 747:124

Feldmann R, Gnedin NY & Kravtsov AV 2012b. Ap. J. 758:127
Field GB, Goldsmith DW & Habing HJ 1969. Ap. J. 155:149
Forbes JC, Krumholz M & Burkert A 2012. Ap. J. 754:48
Forbes JC, Krumholz MR, Burkert A & Dekel A 2014. MNRAS 438:1552
Förster Schreiber NM, Genzel R, Bouché N, et al. 2009. Ap. J. 706:1364
Förster Schreiber NM, Genzel R, Lehnert MD, et al. 2006. Ap. J. 645:1062
Förster Schreiber NM, Übler H, Davies RL, et al. 2019. Ap. J. 875:21
Franco M, Elbaz D, Béthermin M, et al. 2018. Astron. Astrophys. 620:152
Frayer DT, Harris AI, Baker AJ, et al. 2011. Ap. J. 726:22
Freundlich J, Combes F, Tacconi LJ, et al. 2019. Astron. Astrophys. 622:105
Fu J, Guo Q, Kauffmann G & Krumholz MR 2010. MNRAS 409:515
Fu J, Kauffmann G, Li C & Guo Q 2012. MNRAS 424:2701
Fujimoto S, Ouchi M, Shibuya T & Nagai H 2017. Ap. J. 850:83
Galametz M, Madden SC, Galliano F, et al. 2011 Astron. Astrophys. 532:56
Gao Y, Carilli CL, Solomon PM & Vanden Bout PA 2007. Ap. J. Lett. 660:L93
Gao Y & Solomon PM 2004. Ap. J. 606:271
Garcia-Burillo S, Usero A, Alonso-Herrero A, et al. 2012. Astron. Astrophys. 539:8
Genel S, Bouché N, Naab T, Sternberg A & Genzel R 2010. Ap. J. 719:229
Genel S, Genzel R, Bouché N, et al. 2008. Ap. J. 688:789
Genel S, Genzel R, Bouché N, Naab T & Sternberg A 2009. Ap. J. 701:2002
Genel S, Vogelsberger M, Springel V, et al. 2014. MNRAS 445:175
Genzel, R., Lutz, D., Sturm, E. et al. 1998. Ap.J. 498:579
Genzel R, Burkert A, Bouché N, et al. 2008. Ap. J. 687:59
Genzel R, Förster Schreiber NM, Lang P, et al. 2014a. Ap. J. 785:75
Genzel R, Förster Schreiber NM, Rosario, D, et al. 2014b. Ap. J. 796:7
Genzel R, Newman S, Jones T, et al. 2011. Ap. J. 733:101
Genzel R, Tacconi LJ, Combes F, et al. 2012. Ap. J. 746:69
Genzel R, Tacconi LJ, Gracia-Carpio J, et al. 2010. MNRAS 407:2091
Genzel R, Tacconi LJ, Lutz D, et al. 2015. Ap. J. 798:1 (G15)
Giavalisco M, Ferguson HC, Koekemoer AM, et al. 2004. Ap. J. Lett. 600:L93
Gillmon K, Shull JM, Tumlinson J & Danforth C 2006. Ap. J. 636:891
Gilmore G, Wyse RFG & Kuijken K 1989. Annu. Rev. Astron. Astrophys. 27:555
Glover SCO & Clark PC 2012. MNRAS 421:9
Gnedin NY 2000. Ap. J. 542:535
Gnedin NY & Kravtsov AV 2011. Ap. J.728:88
González-López J, Decarli R, Pavesi R, et al. 2019. arXiv:1903.09161
Gracia-Carpio J. 2009. PhD thesis, University of Madrid
Gracia-Carpio J, Garcia-Burillo S, Planesas P, Fuente A & Usero A 2008. Astron. Astrophys. 479:703
Gracia-Carpio J, Sturm E, Hailey-Dunsheath S, et al. 2011. Ap. J. Lett. 728:L7
Grenier IA, Casandjian J-M & Terrier R 2005. Science 307:1292
Grogin NA, Kocevski DD, Faber SM, et al. 2011. Ap. J. Suppl. 197:35
Guimarães R, Petitjean P, de Carvalho RR, et al. 2009. Astron. Astrophys. 508:133
Guo Q & White SDM 2008. MNRAS 384:2
Guo Q, White S, Boylan-Kolchin M, et al. 2011. MNRAS 413:101
Harrison CM, Alexander DM, Mullaney JR, et al. 2014. MNRAS 441:3306

Hatsukade B, Kohno K, Yamaguchi Y, et al. 2018. Publ. Astron. Soc. Japan 70:105
Hayashi M, Tadaki K, Kodama T, et al. 2018. Ap. J. 856:118
Haynes MP, Giovanelli R, Martin AM, et al 2011. Astron. J. 142:170
Heckman TM, Robert C, Leitherer C, Garnett DR & van der Rydt F 1998. Ap. J. 503:646
Heiles C, & Troland TH 2003. Ap. J. Suppl. 145:329
Herrera-Camus R, Bolatto AD, Wolfire MG, et al. 2015. Ap. J. 800:1
Herrera-Camus R, Tacconi L, Genzel R, et al. 2019. Ap. J. 871:37
Heiderman A, Evans NJ II, Allen LE, Huard T & Heyer M 2010. Ap. J. 723:1019
Hickox RC, Mullaney JR, Alexander DM, et al. 2014. Ap. J. 782:9
Hodge JA, Karim A, Smail I, et al. 2013. Ap. J. 768:91
Hodge JA, Smail I, Walter F, et al. 2019. Ap. J. 876:130
Hodge JA, Swinbank AM, Simpson JM, et al. 2016. Ap. J. 833:103
Hopkins PF, Kereš D, Oñorbe J, et al. 2014. MNRAS 445:581
Hu CY, Naab T, Walch S, Glover SCO & Clark PC 2016. MNRAS 458:3258
Huang M-L & Kauffmann G 2014. MNRAS 443:1329
Ilbert O, McCracken HJ, Le Fèvre O, et al. 2013. Astron. Astrophys. 556:551
Ilbert O, Salvato M, Le Floc'h E, et al. 2010. Ap. J. 709:644
Immeli A, Samland M, Gerhard O & Westera P 2004. Astron. Astrophys. 413:547
Israel, FP 1997. Astron. Astrophys. 328:471
Israel FP 2000. Molecular Hydrogen in Space, eds. F. Combes and G Pineau des Forêts. Cambridge: Cambridge University Press, p.293
Ivison RJ, Papadopoulos PP, Smail I, et al. 2011. MNRAS 412:1913
Jansen DJ, Spaans, M Hogerheijde MR & van Dishoeck, EF 1995. Astron. Astrophys. 303:541
Kaasinen M, Scoville N, Walter F, et al. 2019. Ap. J. 880:15
Karim A, Schinnerer E, Martınez-Sansigre A, et al. 2011. Ap. J. 730:61
Kauffmann G, Heckman TM, White SDM, et al. 2003. MNRAS 341:54
Kauffmann G, White SDM & Guiderdoni B 1993. MNRAS 264:201
Kauffmann G, White SDM, Heckman TM, et al. 2004 MNRAS 353:713
Kassin SA, Weiner BJ, Faber SM, et al. 2007. Ap. J. 660:35
Kassin SA, Weiner BJ, Faber SM, et al. 2012. Ap. J. 758:106
Kenney JDP & Young JS 1989. Ap. J. 344:171
Kennicutt RC Jr. 1989. Ap. J. 344:685
Kennicutt RC Jr 1998. Ap. J. 498:541
Kennicutt RC Jr, Calzetti D, Walter F, et al. 2007. Ap. J. 671:333
Kennicutt RC Jr & Evans N 2012. Annu. Rev. Astron. Astrophys. 50:531
Keres D, Yun MS & Young JS 2003. Ap. J. 582:659
Klessen RS, Heitsch F & Mac Low M-M 2000. Ap. J. 535:887
Koekemoer AM, Faber SM, Ferguson HC, et al. 2011. Ap. J. Suppl. 197:36
Könyves V, André P, Men'shchikov A, et al. 2015. Astron. Astrophys. 584:91
Koyama S, Koyama Y, Yamashita T, et al. 2017. Ap. J. 847:137
Kravtsov AV, Berlind AA, Wechsler RH, et al. 2004. Ap. J. 609:35
Kroupa P 2001. MNRAS 322:231
Kruijssen JMD, Schruba A, Chevance M, et al. 2019. Nature 569:519
Krumholz MR, Burkhart B, Forbes JC & Crocker RM 2018. MNRAS 477:2716
Krumholz MR, Dekel A & McKee CF 2012. Ap. J. 745:69

Krumholz MR, Leroy AK & McKee CF 2011. Ap. J. 731:25
Krumholz MR & McKee CF 2005. Ap. J. 630:250
Krumholz MR, McKee CF & Tumlinson J 2008. Ap. J. 689:865
Krumholz M, McKee CF & Tumlinson J 2009. Ap. J. 699:850
Krumholz MR & Tan JC 2007. Ap. J. 654:304
Kuhlen M, Krumholz MR, Madau P, Smith BD & Wise J 2012. Ap. J. 749:36
Lacy JH, Knacke R, Geballe TR & Tokunaga AT 1994. Ap. J. Lett. 428:L69
Lacy JH, Sneden C, Kim H & Jaffe DT 2017. Ap. J. 838:66
Lada CJ, Forbrich J, Lombardi M & Alves JF 2012. Ap. J. 745:190
Lagos CDP, Baugh CM, Lacey CG, et al. 2011. MNRAS 418:1649
Lagos CDP, Bayet E, Baugh CM, et al. 2012. MNRAS 426:2142
Lagos CDP, Crain RA, Schaye J, et al. 2015. MNRAS 452:3815
Lang P, Wuyts S, Somerville R, et al. 2014. Ap. J. 788:11
Larson RB 1981. MNRAS 194:809
Lee MM, Tanaka I, Kawabe R, et al. 2017. Ap. J. 842:55
Lee Y, Snell RL & Dickman RL 1990. Ap. J. 355:536
Lenkic L, Bolatto AD, Förster Schreiber NM, et al. 2019. arXiv:1908.01791
Leroy AK, Bolatto A, Gordon K, et al. 2011. Ap. J. 737:12
Leroy AK, Walter F, Brinks E, et al. 2008. Astron. J. 136:2782
Leroy AK, Walter F, Bigiel F, et al. 2009. Astron. J. 137:4670
Leroy AK, Walter F, Sandstrom K, et al. 2013. Astron. J. 146:19
Li TY, Wechsler RH, Devaraj K & Church SE 2016. Ap. J. 817:169
Lidz, A, Furlanetto SR, Oh SP, et al. 2011. Ap. J. 741:L70
Lilly SJ, Carollo CM, Pipino A, Renzini A & Peng Y 2013. Ap. J. 772:119
Lilly SJ, LeFevre O, Hammer F & Crampton D 1996. Ap. J. 460:1
Lupi A, Bovino S, Capelo PR, Volonteri M & Silk J 2018. MNRAS 474:2884
Lutz D, Poglitsch A, Altieri B, et al. 2011. Astron. Astrophys. 532:90
Mac Low M-M 1999. Ap. J. 524:169
Madau P & Dickinson M 2014. Annu. Rev. Astron. Astrophys. 52:415 (MD14)
Madau P, Ferguson HC, Dickinson ME, et al. 1996. MNRAS 283:1388
Magdis GE, Daddi E, Béthermin M, et al. 2012b. Ap. J. 760:6
Magdis GE, Daddi E, Sargent M, et al. 2012a. Ap. J. 758:9
Magdis GE, Elbaz D, Dickinson M, et al. 2011. Astron. Astrophys. 534:15
Magdis GE, Rigopoulou D, Daddi, E, et al. 2017. Astron. Astrophys. 603:93
Magnelli B, Lutz D, Saintonge A, et al. 2014. Astron. Astrophys. 561:86
Magnelli B, Saintonge A, Lutz D, et al. 2012. Astron. Astrophys. 548:22
Marinacci F, Grand RJJ, Pakmor R, et al. 2017. MNRAS 466:3859
Martin CL, Scannapieco E, Ellison SL, et al. 2010. Ap. J. 721:174
Mashian N, Sternberg A & Loeb A 2015. Jour. Cosm. Astropart. Phys. 11:28
McKee CF & Ostriker EC 2007. Annu. Rev. Astron. Astrophys. 45:565
McKee CF & Krumholz M 2010. Ap. J. 709:308
Meidt SE, Leroy AK, Rosolowsky E, et al. 2018. Ap. J. 854:100
Michałowski MJ, Dunlop JS, Cirasuolo M, et al. 2012. Astron. Astrophys. 541:12
Mihos JC & Hernquist L 1996. Ap. J. 464:641
Miville-Deschenes MA, Murray N & Lee EJ 2017. Ap. J. 834:57
Mo HJ, Mao S & White SDM 1998. MNRAS 295:319 (MMW)

Mok A., Wilson CD, Golding J, et al. 2016. MNRAS 456:4384
Momcheva IG, Brammer GB, van Dokkum PG, et al. 2016. Ap. J. Suppl. 225:27
Mortier AMJ, Serjeant S, Dunlop JS, et al. 2005. MNRAS 363:563
Moster BP, Naab T & White SDM 2013. MNRAS 428:3121
Mouschovias TC 1976. Ap. J. 207:141
Mullaney JR, Pannella M, Daddi E, et al. 2012. MNRAS 419:95
Murray N, Quataert E & Thompson TA 2005. Ap. J. 618:569
Narayanan D, Krumholz M, Ostriker EC & Hernquist L 2011. MNRAS 418:664
Narayanan D, Krumholz M, Ostriker EC & Hernquist L 2012. MNRAS 421:3127
Neistein E & Dekel A 2008. MNRAS 388:1792
Nelson D, Pillepich A, Springel V, et al. 2019. arXiv:1902.05554
Newman SF, Genzel R, Förster Schreiber NM, et al. 2013. Ap. J. 767:104
Nickerson S, Teyssier R & Rosdahl J 2018. MNRAS 479:3206
Nickerson S, Teyssier R & Rosdahl J 2019. MNRAS 484:1238
Noble AG, McDonald M, Muzzin A, et al. 2017. Ap. J. Lett. 842:L21
Noeske KG, Weiner BJ, Faber SM, et al. 2007. Ap. J. Lett. 660:L43
Noguchi M 1999. Ap. J. 514:77
Nordon R, Lutz D, Genzel R, et al. 2012. Ap. J. 745:182
Nordon R & Sternberg A 2016. MNRAS 462:2804
Noterdaeme P, Petitjean P, Carithers WC, et al. 2012. Astron. Astrophys. 547:1
Obreschkow D, Croton D, De Lucia G, Khochfar S & Rawlings S 2009. Ap. J. 698:1467
Obreschkow D & Rawlings S 2009a. MNRAS 394:1857
Obreschkow D & Rawlings S 2009b. Ap. J. Lett. 696:L129
Oesch PA, Brammer G, van Dokkum PG, et al. 2016. Ap. J. 819:129
Oliver SJ, Bock J, Altieri B, et al. 2012. MNRAS 424:1614
Oppenheimer BD, Davé R, Kereš D, et al. 2010. MNRAS 406:2325
Ostriker EC, McKee CF & Leroy AK 2010. Ap. J. 721:975
Ostriker EC & Shetty R 2011. Ap. J. 731:41
Pannella M, Carilli CL, Daddi E, et al. 2009. Ap. J. Lett. 698:L116
Papadopoulos PP, Bisbas TG & Zhang Z-Y 2018. MNRAS 478:1716
Parmar PS, Lacy JH & Achtermann JM 1991. Ap. J. Lett. 372:L25
Patil P, Nyland K, Lacy M, et al. 2019. Ap. J. 871:109
Pavesi R, Sharon CE, Riechers DA, et al. 2018. Ap. J. 864:49
Peng Y, Lilly SJ, Kovač K, et al. 2010. Ap. J. 721:193
Peng Y & Maiolino R. 2014. MNRAS 443:3643
Peng Y, Maiolino R & Cochrane R 2015. Nature 521:192
Péroux C, Dessauges-Zavadsky M, D'Odorico S, Sun KT & McMahon RG 2005. MNRAS 363:479
Pettini M & Pagel BEJ 2004. MNRAS 348:59
Pillepich A, Nelson D, Springel V, et al. 2019. arXiv:1902.05553
Popping G, Caputi KI, Trager SC, et al. 2015. MNRAS 454:2258
Popping G, Pillepich A, Somerville RS, et al. 2019 arXiv:1903.09158
Popping G, Somerville RS & Trager SC 2014. MNRAS 442:2398
Popping G, van Kampen E, Decarli R, et al. 2016. MNRAS 461:93
Prochaska JX & Wolfe AM 2009. Ap. J. 696:1543
Rachford BL, Snow TP, Destree JD, et al. 2009. Ap. J. Suppl. 180:125

Rachford BL, Snow TP, Tumlinson J, et al. 2002. Ap. J. 577:221
Rathaus B & Sternberg A 2016. MNRAS 458:3168
Rao SM, Turnshek DA & Nestor DB 2006. Ap. J. 636:610
Rees MJ & Ostriker JP 1977. MNRAS 179:541
Rémy-Ruyer A, Madden SC, Galliano F, et al. 2014. Astron. Astrophys. 563:31
Renzini A & Peng Y 2015. Ap. J. 801:29
Rhee J, Lah P, Briggs FH, et al. 2018. MNRAS 473:1879
Richter MJ, Graham JR, Wright GS, Kelly DM & Lacy JH 1995. Ap. J. Lett. 449:L83
Riechers DA, Carilli CL, Walter F & Momjian E 2010. Ap. J. Lett. 724:L153
Riechers DA, Pavesi R, Sharon CE, et al. 2019. Ap. J. 872:7
Rieke GH, Lebofsky MJ, Thompson RI, Low FJ & Tokunaga AT 1980. Ap. J. 238:24
Rieke GH, Loken K, Rieke MJ & Tamblyn P 1993. Ap. J. 412:99
Righi, M, Hernandez-Monteagudo, C & Sunyaev 2008. Astron. Astrophys. 489:489
Robertson BE & Kravtsov AV 2008. Ap. J. 680:1083
Rodighiero G, Cimatti A, Gruppioni C, et al. 2010. Astron. Astrophys. 518:25
Rodighiero G, Daddi E, Baronchelli I, et al. 2011. Ap. J. 739:40
Rodighiero G, Brusa, M, Daddi E, et al. 2015. Ap. J. 800:10
Romanowsky AJ & Fall SM 2012. Ap. J. Suppl. 203:17
Rowan-Robinson M & Crawford J 1989. MNRAS 238:523
Rudnick G, Hodge J, Walter F, et al. 2017. Ap. J. 849:27
Saintonge A, Catinella B, Cortese L, et al. 2016. MNRAS 462:1749
Saintonge A, Catinella B, Tacconi LJ, et al. 2017. Ap. J. Suppl. 233:22
Saintonge A, Kauffmann G, Kramer C, et al. 2011a. MNRAS 415:32
Saintonge A, Kauffmann G, Wang J, et al. 2011b. MNRAS 415:61
Saintonge A, Lutz D, Genzel R, et al. 2013. Ap. J. 778:2
Sanders DB & Mirabel IF 1996. Annu. Rev. Astron. Astrophys. 34:749
Sanders DB, Soifer BT, Elias JH, et al. 1988. Ap. J. 325:74
Santini P, Maiolino R, Magnelli B, et al. 2014. Astron. Astrophys. 562:30
Sargent MT, Béthermin M, Daddi E, et al. 2014. Ap. J. 793:19
Scalo J & Elmegreen BG 2004. Annu. Rev. Astron. Astrophys. 42:275
Schaye J, 2004. Ap. J. 609:667
Schaye J, Crain RA, Bower RG, et al. 2015. MNRAS 446:521
Schiminovich D, Wyder TK, Martin DC, et al. 2007. Ap. J. Suppl. 173:315
Schinnerer E, Groves B, Sargent MT, et al. 2016. Ap. J. 833:112
Schreiber C, Pannella M, Elbaz D, et al. 2015. Astron. Astrophys. 757:74
Schruba A, Kruijssen JMD & Leroy AK 2019. in prep?
Schruba A, Leroy AK, Walter F, et al. 2011. Astron. J. 142:37
Scoville NZ, Aussel H, Sheth K, et al. 2014. Ap. J. 783:84
Scoville NZ, Lee N, van den Bout P, et al. 2017. Ap. J. 837:150 (S17)
Scoville NZ, Sheth K, Aussel H, et al. 2016. Ap. J. 820:83
Scoville NZ, Yun MS & Bryant PM 1997. Ap. J. 484:702
Shapiro KL, Genzel R, Förster Schreiber NM, et al. 2008. Ap. J. 682:23
Sharon CE, Tagore AS, Baker AJ, et al. 2019. Ap. J. 879:52
Shu FH, Adams FC & Lizano S 1987. Annu. Rev.Astron. Astrophys. 25:23
Shull JM, Tumlinson J, Jenkins EB, et al. 2000. Ap. J. 538:73
Silk J 1997. Ap. J. 481:703

Silverman JD, Daddi E, Rodighiero G, et al. 2015. Ap. J. 812:23
Silverman JD, Rujopakarn W, Daddi E, et al. 2018. Ap. J. 867:92
Simons RC, Kassin SA, Weiner BJ, et al. 2017. Ap. J. 843:46
Skelton RE, Whitaker KE, Momcheva IG, et al. 2014. Ap. J. Suppl. 214:24
Smail I, Ivison RJ & Blain AW 1997. Ap. J. Lett. 490:L5
Solomon PM, Downes D, Radford SJE & Barrett JW 1997. Ap. J. 478:144
Solomon PM, Rivolo AR, Barrett J & Yahil A 1987. Ap. J. 319:730
Somerville RS & Davé, R. 2015. Annu. Rev. Astron. Astrophys. 53:51
Speagle JS, Steinhardt CL, Capak PL & Silverman JD 2014. Ap. J. Suppl. 214:15 (S14)
Spilker J, Bezanson R, Barišić I, et al. 2018. Ap. J. 860:103
Stacey GJ, Hailey-Dunsheath S, Ferkinhoff C, et al. 2010. Ap. J. 724:957
Stanimirović S, Murray CE, Lee MY, Heiles C & Miller J 2014. Ap. J. 793:132
Steidel CC, Giavalisco M, Dickinson ME & Adelberger KL 1996. Astron. J. 112:352
Sternberg A 1988. Ap. J. 332:400
Sternberg A & Dalgarno A 1989. Ap. J. 338:197
Sternberg A & Dalgarno A 1995. Ap. J. Suppl. 99:565
Sternberg A, Le Petit F, Roueff E & Le Bourlot J 2014. Ap. J. 790:10
Stone JM, Ostriker EC & Gammie CF 1998. Ap. J. Lett. 508:L99
Stott JP, Swinbank AM, Johnson HL, et al. 2016. MNRAS 457:1888
Strong AW & Mattox JR 1996. Astron. Astrophys. 308:L21
Suess KA, Bezanson R, Spilker JS, et al. 2017. Ap. J. Lett. 846:L14
Sun J, Leroy AK, Schruba A, et al. 2018. Ap. J. 860:172
Swinbank AM, Harrison CM, Trayford J, et al. 2017. MNRAS 467:3140
Tacchella S, Carollo CM, Renzini A, et al. 2015. Science 348:314
Tacchella S, Dekel A, Carollo CM, et al. 2016. MNRAS 457:2790
Tacconi LJ, Genzel R, Neri R, et al. 2010. Nature 463:781
Tacconi LJ, Genzel R, Saintonge A, et al. 2018. Ap. J. 853:179 (T18)
Tacconi LJ, Genzel R, Smail I, et al. 2008. Ap. J. 680:246
Tacconi LJ, Neri R, Chapman SC, et al. 2006. Ap. J. 640:228
Tacconi, LJ, Neri R, Genzel R, et al. 2013. Ap. J. 768:74
Tadaki K, Genzel R, Kodama T, et al. 2017. Ap. J. 834:135
Tadaki K, Kodama T, Hayashi M, et al. 2019. Publ. Astron. Soc. Japan 71:40
Thompson R, Nagamine K, Jaacks J & Choi J-H, 2014. Ap. J. 780:145
Tomassetti M, Porciani C, Romano-Diaz E, Ludlow AD & Papadopoulos PP 2014. MNRAS 445:124
Tomassetti M, Porciani C, Romano-Díaz E & Ludlow, AD 2015. MNRAS 446:3330
Toomre A 1964. Ap. J., 139:1217
Trujillo I, Feulner G, Goranova Y, et al. 2006. MNRAS 373:36
Tumlinson J, Shull JM, Rachford BL, et al. 2002. Ap. J. 566:857
Übler H, Genzel R, Tacconi LJ, et al. 2018. Ap. J. Lett. 854:L24
Übler H, Genzel R, Wisnioski E, et al. 2019. Ap. J. 880:48
Utomo D, Bolatto AD, Wong T, et al. 2017. Ap. J. 849:26
Valentino F, Magdis GE, Daddi E, et al. 2018. Ap. J. 869:27
Vallini L, Gruppioni C, Pozzi F, Vignali C & Zamorani G 2016. MNRAS 456:40
Valtchanov I, Virdee J, Ivison RJ, et al. 2011. MNRAS 415:3473
Vanden Bout PA, Solomon PM & Maddalena RJ 2004. Ap. J. Lett. 614:L97

van Dokkum PG, Franx M, Kriek M, et al. 2008. Ap. J. Lett. 677:L5
Walter F, Decarli R, Aravena M, et al. 2016. Ap. J. 833:67
Warren SR, Skillman, ED, Stilp AM, et al. 2012. Ap. J. 757:84
Weiss A, Downes D, Walter F & Henkel C 2007. From Z-Machines to ALMA: (Sub)Millimeter Spectroscopy of Galaxies, eds. AJ Baker, J Glenn, AI Harris, JG Mangum, MS Yun (San Francisco, CA: ASP), 25
Weiß A, Kovács A., Coppin K, et al. 2009. Ap. J. 707:1201
Wetzel AR, Hopkins PF, Kim J, et al. 2016. Ap. J. 823:23
Whitaker KE, Franx M, Leja J, et al. 2014. Ap. J. 795:104
Whitaker KE, van Dokkum PG, Brammer G & Franx M 2012. Ap. J. Lett. 754:29L
White SDM & Frenk CS 1991. Ap. J. 379:52
White SDM & Rees MJ 1978. MNRAS 183:341
Wiklind T, Conselice CJ, Dahlen T, et al. 2014. Ap. J. 785:111
Wiklind T, Ferguson HC, Guo Y, et al. 2019. Ap. J. 878:83
Williams RE, Blacker B, Dickinson M, et al. 1996. Astron. J. 112:1335
Williams RJ, Quadri RF, Franx M, van Dokkum P & Labbé I 2009. Ap. J. 691:1879
Wilson CD 1995. Ap. J. Lett. 448:L97
Wisnioski E, Förster Schreiber NM, Wuyts S, et al. 2015. Ap. J. 799:209
Wisnioski E, Förster Schreiber NM, Fossati M, et al. 2019. Ap. J. submitted
Wolfire MG, Hollenbach D & McKee CF 2010. Ap. J. 716:1191
Wolfire MG, McKee CF, Hollenbach D & Tielens AGGM 2003. Ap. J. 587:278
Wong T & Blitz L 2002. Ap. J. 569:157
Wuyts S, Förster Schreiber NM, Lutz D, et al. 2011a. Ap. J. 738:106
Wuyts S, Förster Schreiber NM, van der Wel A, et al. 2011b. Ap. J. 742:96
Xie L, De Lucia G, Hirschmann M, Fontanot F & Zoldan A 2017. MNRAS 469:968
Young JS & Scoville NZ 1991. Annu. Rev. Astron. Astrophys. 29:581
Zavala JA, Casey CM, da Cunha E, et al. 2018. Ap. J. 869:71
Zolotov A, Dekel A, Mandelker N, et al. 2015. MNRAS 450:2327
Zuckerman B & Evans NJ 1974. Ap. J. 192:149
Zwaan MA, Meyer MJ, Staveley-Smith L & Webster RL 2005. MNRAS 359:30

# 8. Tables

**Table 1 – Surveys of Molecular Gas and Dust for Galaxy Evolution: Sources Used for Gas Scaling Relations**

| Survey | References | Method | Sources | Redshift Range |
|---|---|---|---|---|
| xCOLDGASS[a] | Saintonge+2011, 2016, 2017 | CO | 293 detections 1 stack (non-det) | z~0 |
| CALIFA-EDGE | Bolatto+2017, Utomo+2017 | CO | 100 detections included | z~0 |
| LIRGs/ULIRGs | Armus+2009, Gao & Solomon 2004, Garcia-Burillo+2012, Gracia-Carpio+2008, 2009 | CO | 90 | z~0 |
| CO- FIR SED cross calib. | Bertemes+2018 | CO, SED dust | 78 CO 78 FIR SED | z~0 |
| EGNOG | Bauermeister+2013 | CO | 31 | z~0.05-0.5 |
| z~0.5 ULIRGS | Combes+2011, 2013 | CO | 32 | z~0.2-0.9 |
| PHIBSS [a] | Tacconi+2013, T18, G15, Freundlich+2019 | CO | 131 | z~0.5-2.5 |
| ASPECS | Decarli+2016, 2019 | CO | 22 | z~1-3.5 |
| Other high-z CO (PdBI, ALMA, NOEMA) | Daddi+2010a, G15, Herrera-Camus+2019, Magdis+2012b, 2017, Magnelli+2012, T18, Spilker+2018, Suess+2017 Übler+2018,Valentino+2018 | CO | 35 | z~0.5-3.2 |
| z~1 Starbursts | Silverman+2015, 2018 | CO | 12 | z~1.5 |
| SMGs[b] | Bothwell+2013, Tacconi+2006, 2008 | CO | 18 | z~1.2-4.0 |
| High-z MS lensed SFGs | Saintonge+2013 | CO | 7 | z~1.4-2.7 |

| Survey | Reference | Data Type | Number | Redshift |
|---|---|---|---|---|
| COSMOS CO-dust cross calibration | Kaasinen+2019 | CO, 1mm dust | 8 CO  12 1mm dust | z~1.7-2.9 |
| COSMOS PEP | Béthermin+2015 | FIR SED | 15 stacks | z~0.75-3.8 |
| Herschel PEP + HerMES | Magnelli+2014, Berta+2016 | FIR SED | 510 stacks | z~0.12-2.0 |
| Herschel PEP + HerMES | Santini+2014 | FIR SED | 121 stacks | z~0.12-2.0 |
| PHIBSS | T18, PHIBSS in prep. | 1mm dust[c] | 7 | z~1.1-2.3 |
| 3mmALMA Continuum | Zavala+2018 | 3mm dust | 20 | 1.0-3.7 |
| Scoville COSMOS | Scoville+2016 | 1mm dust | 51 Sources  21 Stacks | 0.9-2.8 |
| DANCING | Fujimoto+2017 | 1mm dust | 102 | 0.6-5.3 |
| ASAGAO | Hatsukade+2018 | 1mm dust | 23 | z~1.2-4.5 |
| GOODS-S | Franco+2018, Cowie+2018 | 1mm dust | 68 | z~0.1-5.5 |
| UDF | Dunlop+2017 | 1mm dust | 12 | z~0.7-3 |
| Tadaki 3D-HST Massive | Tadaki+2019, in prep | 1mm dust | 95 | z~2-2.5 |
| Wiklind CANDELS GOODS-S | Wiklind+2019 | 1mm dust | 25 | z~0.5-4.5 |
| Other ALMA | Aravena+2019, Barro+2016, Decarli+2016, Hodge+2016, 2019, Magdis+2017, Patil+2019, Schinnerer+16, Sharon+2019, G15, T18 | 1mm dust | 2+5+5+1+6+7+6+2=34 | z~2.3-4.8 |

Notes to Table 1:
[a]Here we use a 3.9σ significance cut, which results in fewer sources included than in T18
[b]Not a complete list, but the same sample as used in T18

[c]Here we refer as "1mm dust" all those studies using the single ~1mm continuum measurement to derive molecular gas masses, by the method described in S16. The observed wavelength varies from study to study, and we refer the reader to the references listed for more details.

**Table 2a.** Fit parameters for $t_{depl}$ (equation (4)) obtained from error-weighted, multi-parameter regression:
$\log t_{depl}(Gyr) = A + B \times \log(1+z) + C \times \log(sSFR/sSFR(MS,z,M_*)) + D \times (\log M_* - 10.7)$

| Data | parameter | N | $\chi^2_r$ | A | B | C | D |
|---|---|---|---|---|---|---|---|
| all (error weighted) | $t_{depl}$ (Gyr) S14 | 2052 | 0.85 | $+0.21_{0.1}$ | $-0.98_{0.1}$ | $-0.49_{0.03}$ | $+0.03_{0.04}$ |

**Table 2b.** Fit parameters for $\mu_{molgas}$ obtained from error-weighted, multi-parameter regression:
$\log M_{molgas}/M_* = A + B \times (\log(1+z) - F)^2 + C \times \log(sSFR/sSFR(MS,z,M_*)) + D \times (\log M_* - 10.7)$

| Data | parameter | N | $\chi^2_r$ | A | B | F | C | D |
|---|---|---|---|---|---|---|---|---|
| all (error weighted) | $\mu_{molgas}$ $=M_{molgas}/M_*$ S14 | 2052 | 0.75 | $+0.06_{0.2}$ | $-3.33_{0.2}$ | $+0.65_{0.05}$ | $+0.51_{0.03}$ | $-0.41_{0.03}$ |

# 9. List of Definitions

1. GMC: Giant Molecular Cloud with typical properties M~$10^4$…$10^{6.5}$ M$_\odot$), n(H$_2$)~$10^2$…$10^5$ cm$^{-3}$ and T$_{gas}$~10-40 K
2. SFR: star formation rate in M$_\odot$ yr$^{-1}$
3. SFGs, ETGs: star forming galaxies, early-type galaxies
4. Q: The Toomre stability parameter, which is <0.67 (1) for a thick (thin) gaseous disk unstable against gravitational collapse, and >1 if there is also a stellar disk
5. MS, the star formation main sequence – The tight correlation between stellar mass and SFR for ~90% of SFGs
6. Specific star formation rate, sSFR = SFR/M*, where M* is the stellar mass. sSFR normally has units of Gyr$^{-1}$
7. $\alpha_{CO}$, the CO-to-H$_2$ conversion factor
8. Molecular gas depletion time scale, $t_{depl} = M_{molgas}/SFR$
9. gas to stellar mass ratio, $\mu_{molgas} = M_{molgas}/M*$
10. molecular gas fraction, $f_{molgas} = M_{molgas}/(M_{molgas}+M*)$
11. Main sequence offset, $\delta MS=log(SFR/SFR(MS))$, the logarithmic offset of a galaxy at constant mass and redshift from the MS line. In this review we adopt the MS parameterization of Speagle et al. 2014
12. Outflow mass-loading factor, $\eta = \dot{M}_{out,wind}/SFR$
13. molecular cloud life-time $t_{GMC}$
14. free-fall time $t_{ff}$ and star formation efficiency per free-fall time $\varepsilon_{ff}$
15. galactic dynamical ($t_{dyn}$) and Toomre time ($t_T$)
16. growth time of the dark matter halo, $t_{DM}$, of the molecular gas mass, $t_{acc}$, and of the stellar mass $t_*$
17. Hubble time $H(z)^{-1}$
18. time between major ($t_{major\,merger}$) and minor ($t_{minor\,merger}$) mergers
19. time for replenishment of the molecular gas reservoir $t_{reservoir-repl}=M_{molgas}/(dM_{molgas}/dt)$
20. radial transport time $t_{dv}$ due to dynamical friction, torques and viscosity
21. recyclng time $t_{rec}$~0.5-1 Gyr, for gas ejected by (stellar) feedback to return back to the galaxy, in effect increasing the gas accretion rate from the CGM above the level of cosmic accretion, especially at z~0.5-1.5

# 10. Sidebars

1) *Quenching.* The redshift independent maximum stellar mass requires a strongly mass dependent 'quenching' process that terminates further mass growth and causes galaxies to transition from the star forming MS population to the red sequence of passive galaxies (e.g. Kauffmann et al. 2004; Faber et al. 2007). It is not yet clear whether this quenching process originates at the scale of the dark matter halos, by transitions to hot, slow cooling quasi-isothermal gaseous envelopes above halo masses M$_{crit}$~$10^{12}$ M$_\odot$ (Rees & Ostriker 1977; Dekel & Birnboim 2006), or whether galaxy internal processes are the main quenching drivers. For field galaxies, the probability of being quenched correlates most strongly with the central stellar surface density, Sersic index and bulge mass (Kauffmann et al. 2003; Bell et al. 2012; Lang et al. 2014). This suggests that high velocity outflows from the central black holes drive out gas from the galaxy, and create hot buoyant CGM atmospheres that prevent further accretion and quench star formation (Croton et al. 2006; Bower et al. 2017; Nelson et al. 2019; Faerman et al. 2017). Scaling relations of ionized outflows at z~1-2.5 support this proposal (e.g. Genzel et al. 2014a; Harrison et al. 2014; Brusa et al. 2015; Förster Schreiber et al. 2019). In galaxy clusters and dense groups, starvation and harassment on large scales are additional mechanisms, even for lower mass galaxies (Peng et al. 2010, 2015).

Blue SFGs on and above the MS, and red passive early type galaxies (ETGs) below the MS are well separated in UV/optical color-color space (Williams et al. 2009; Renzini & Peng 2015). *SFRs* are uncertain below the main sequence, however, where mid/far-IR detections are rare or non-existent (e.g. Wuyts et al. 2011a). Eales et al. (2015, 2017) have shown that far-infrared based *SFRs* in the near infrared selected, volume-limited Herschel Reference Survey in the local Universe (Boselli et al. 2010) exhibit a much more gradual transition from active star formers to passive galaxies. The sharp dichotomy seen in optically or UV-selected samples could thus be somewhat misleading, perhaps due to variations in dustiness and errors in *SFRs* based on spectral energy distributions.

*2) **The 'Toomre-Jeans' Instability Cascade.*** In gas-rich, rotating disk galaxies with disk scale length $R_d$, circular velocity $v_c$, local velocity dispersion $\sigma_0$, and gas surface density $\Sigma_{gas}$ differential rotation stabilizes cloud complexes against fragmentation on scales larger than the Toomre length ($\lambda_T = (\pi G \Sigma_{gas} R_d^2)/v_c^2 \simeq R_d \times f_{gas}$; Toomre 1964). Between this 'upper' stability scale and the pressure stability scale, the Jeans length $\lambda_{Jeans} \sim \sigma_0^2/G\Sigma_{gas}$, all intermediate scales are unstable and the most rapidly growing mode has a wavelength of 2 $\lambda_{Jeans}$ (Toomre 1964; Binney & Tremaine 2008; Dekel et al. 2009; Escala 2011). In finite thickness, gas rich and globally Toomre-unstable disks ($Q = \dfrac{\sqrt{2}\sigma_0 \times v_c}{\pi G \Sigma_{gas} \times R_d} \leq Q_{crit} \sim 0.7$) fragmentation and cloud collapse do not occur on the local free-fall time scale but on the Toomre time, $t_T = \dfrac{R_d}{v_c} \times \dfrac{\left[(2Q^2)^{-1} - 1\right]^{-1/2}}{\sqrt{2}} \xrightarrow{Q < Q_{crit} \sim 0.7} t_{dyn} \times Q$, where $t_{dyn}=R_d/v_c$ is the galactic dynamical time (Behrendt et al. 2015; Burkert et al. 2019; Elmegreen 1997; Silk 1997; Krumholz et al. 2012).

*3) **Recycling.*** For momentum driven winds, the outflow mass-loading factor, $\eta = \dot{M}_{out,wind}/SFR$ (and the ratio of outflow to circular velocity) should correlate with galaxy mass $M$ as $\eta \sim M^{-1/3}$, Murray et al. 2005; Oppenheimer et al. 2010), or with $\eta \sim M^{-2/3}$ for energy driven winds (Faucher-Giguère & Quataert 2012). Small galaxies (at high-z) would form inefficiently and lose much of the incoming gas to the CGM, in agreement with abundance matching results (e.g. Behroozi et al. 2010; Moster et al. 2013). If the outflow velocities are not sufficient to fully expel the gas from the halo, some could return as fountains at later times, when the galaxy is more massive, resulting in accretion rates and gas fractions above $\eta$=const models (Davé et al. 2011a; Dekel & Mandelker 2014). The 'recycling time' $t_{rec}$ scales inversely with mass, such that for galaxies above *log $(M_*/M_\odot) \geq$* 10.5 the recycling time is ~0.5-1 Gyr (Oppenheimer et al. 2010), so that gas ejected at $z$~2 returns at $z$~1.5 and could substantially increase gas accretion rates, star formation rates and gas reservoir masses. Recycling processes are probably only well captured in hydro-simulations with strong winds, which do well on the peak reservoir redshift in Figure 1 and 7, and much better for the high mass gas fractions.